\documentclass[journal]{IEEEtran}

\usepackage{amsmath,amsfonts}
\usepackage{algorithmic}
\usepackage{algorithm}
\usepackage{array}
\usepackage[caption=false,font=normalsize,labelfont=sf,textfont=sf]{subfig}
\usepackage{textcomp}
\usepackage{stfloats}
\usepackage{url}
\usepackage{verbatim}
\usepackage{graphicx}
\usepackage{cite}
\def\BibTeX{{\rm B\kern-.05em{\sc i\kern-.025em b}\kern-.08em
    T\kern-.1667em\lower.7ex\hbox{E}\kern-.125emX}}
\usepackage{balance}

\usepackage{xspace}
\usepackage{soul}

\newcommand{\dg}{$^\circ$\xspace}
\newcommand{\um}{$\mu$m\xspace}

\newcommand{\mms}{mm$^2$\xspace}

\begin{document}

\title{Scalable switched slab coupler based optical phased array on silicon nitride}

\author{P. Muñoz, D. Pastor, L.A. Bru, G. Micó, J. Benítez, D.J. Goodwill and E. Bernier
\thanks{Manuscript created October, 2021; G. Micó, L.A. Bru, J. Benítez, D. Pastor and P. Muñoz are with the Photonic Research Labs, Institute for Telecommunication and Multimedia Applications and UPVfab, Universitat Politècnica de València.}
\thanks{D.J. Goodwill and E. Bernier are with Huawei Canada Co., Ltd. - 303 Terry Fox Drive, Ottawa, ON K2K 3J1, Canada.}
\thanks{Corresponding author P. Muñoz, pascual.munoz@upv.es}
}

\maketitle

\begin{abstract}
A two-dimensional optical-phased array is demonstrated by using a multiple-input star coupler, compatible with FMCW LiDAR. Previous approach using a single-input design achieves two-dimensional beam-steering by relying on a tunable laser source, taking advantage of grating coupler radiation angle wavelength dependance and variation of the waveguide refractive index. While implementing a convenient way to distribute power in a single step, star coupler architecture is inefficient in terms of employed waveguide length and thus, optical loss and footprint. Multi-input approach partially alleviates this by condensing several single-input devices into one, permitting to reduce the footprint proportionally to the employed number of inputs. We fabricated in silicon nitride technology a proof of concept steerer with beam waist 0.36\dg$\times$0.175\dg  addressing a field of view of 15\dg$\times$2.8\dg. A new design iteration is also reported with and 0.24\dg$\times$0.16\dg beam waist and 15\dg$\times$11.2\dg field of view. Implications of this optical-phased array chips from a LiDAR system perspective are also presented. 
\end{abstract}

\begin{IEEEkeywords}
photonic integrated circuit, optical phase array, LiDAR, FM-CW, silicon nitride
\end{IEEEkeywords}

\IEEEpeerreviewmaketitle

\section{Introduction}

\IEEEPARstart{C}{ompact} and functional light detection and ranging (LiDAR) system requires a working combination of laser, beam-steering system, detector and digital signal processing (DSP) \cite{Zhu:20}. Hybrid integration in photonic integrated circuits (PIC) is a promising platform as it permits to integrate active devices such as lasers and photo-detectors, as well as integrated electronic circuits, along with optical waveguides. In this context, solid-state optical phased arrays (OPA) beam-steering have been extensively demonstrated to have advantages over mechanical scanners, including chip-scale dimensions, reliability, low-cost manufacturing and adaptive field-of-view and sweep rate \cite{Guo:21}. On the other hand, there are two main system level approaches to implement LiDAR: time-of-flight (ToF) and frequency-modulated continuous wave (FMCW). The former requires the use of multi-spatial-mode lasers, incompatible with single-mode integrated waveguides. In FMCW, single spatial mode lasers are employed and, more specifically at $\lambda=1550$ nm, has advantages in terms of signal isolation, minimizing interference with other LiDAR, presenting better rejection to sunlight and achieving longer ranges when compared to ToF.

In this paper we present and experimentally demonstrate a OPA PIC architecture compatible with FMCW. The architecture can randomly address a scene, by combining the operation of a tunable laser and a switch matrix selector. The fundamental OPA block is based on a slab coupler and a waveguide array equipped with grating couplers (SC-OPA), similar to the one proposed in \cite{Van:11}. Nonetheless, our SC-OPA uses a plurality of inputs to the slab coupler, that can be selected by a switch matrix. This results into a SC-OPA block more compact, for a given wavelength tuning range, than the one using a single input. The reduced footprint is key since several SC-OPAs are required on the same PIC, each covering a different part of the scene, as it will be detailed later. Hence, an additional switching stage is employed to address different SC-OPA blocks.

In the literature, several implementation of solid-state OPAs can be found, that resort to tunable lasers and switch matrices. The beam steering dimension exploiting wavelength tuning and diffraction effect on gratings is typically 0.08\dg/nm on silicon nitride and 0.14\dg/nm on silicon on insulator \cite{doerr2010,ito2020}. Other implementations make use of strong dispersive photonic crystals, surrounding the grating structure, and increase the angular dispersion up to 1\dg/nm performing 30\dg steering range for a modest 20~nm laser wavelength tuning range \cite{ito2020}.

Other groups have reported switched architectures applied to a free-space propagation region \cite{Doerr2021}, but in this case the diffraction region is large area focusing grating, instead of an OPA. The underlying input switching on this implementation allows to control horizontal scanning, while in the switched SC-OPA we propose, the waveguide array and tunable laser do also play a role. Furthermore, the architecture in \cite{Doerr2021} does not seem scalable to a large number of steering directions.

The same wavelength tuning controlled double H and V sweep concept has been proposed also in \cite{Dostart2020}. In this case, however, a serpentine waveguide structure with embedded grating couplers is presented. While it results into a very compact layout, it comes with a series of design limitations. The grating coupler separation is then limited by the serpentine waveguide bend radius, and therefore the maximum achievable horizontal field of view is limited. Furthermore,  due to the fact the signal must sequentially pass through the entire structure, the design of the gratings is critical as they need to be very weak, thus low perturbation structures are required.

The footprint and architecture scaling when the number of radiating elements in the OPA increases are root concerns of the different proposals. Bogaerts et. \cite{Bogaerts2021}   present a systematic analysis of footprint and scaling where pure structures such as serpentine, tree structure or slab couplers can be conveniently combined to reduce their size while maintaining the performance. In \cite{Bogaerts2021} an approximation is also presented in which a set of evenly spaced OPA units of smaller size are combined by means of a distribution network without phase changes (minimum size) to achieve a greater total equivalent area. However this comes at the expense of some penalties in terms of main to secondary lobe ratio, and the emission directions allowed are discrete.

Our paper is structured as follows. Next section presents the rationale behind our work, from the specification to the technology selection and proposed architecture. In section~\ref{sec:dsgnfab}, two of our multi-input SC-OPA designs are elaborated. Section~\ref{sec:tech} provides the results of the technology validation and building block performance. Next, section~\ref{sec:test} reports on the experimental setup, methods and test results of the two aforementioned designs. The connection between the PIC and LiDAR system level implications is developed in section~\ref{sec:system}. Finally the conclusion is given in section~\ref{sec:concl}.

\section{\label{sec:spec}Specification, technology and architecture}

\subsection{Specification}
The purpose of our research was to determine the best combination of a scalable chip architecture and photonic integration technology, incorporating an OPA, that ultimately would lead to meet a set of specifications. The sought OPA scanner should cover a 2D scene spanning a field of view (FOV) of $\pm$15\dg in both the horizontal (H) and vertical (V) directions, with a resolution of 0.1\dg. Additionally, the scanner should feature random access to parts of the scene, with the lower power consuming approach. From a system perspective, the resulting architecture should be compatible with FMCW LiDAR. Last but not least, when resorting to off-chip devices, they should be state of the art, and co-integrable with the OPA in the mid-term.

One on-chip key block, as in many other proposals in the literature, is the grating coupler (GC) as radiating element. These radiate light on different directions for different wavelengths. Thus, with a tunable laser (TL), one of the two dimensions in the FOV, in our case V, can be addressed. Consequently, the phased-array is constructed, in the H dimension, by placing a number of ideally identical GCs, $N_{GC}$, spaced $d_{GC}$ from each other. The scan in H can then resorts to different mechanisms that have implications in the power distribution network to the GCs. We soon discarded architectures based on meander or serpentine waveguides such as \cite{Dostart2020} mainly for two reasons: footprint when increasing $N_{GC}$ and beam quality in H, due to the difficulties in precisely setting in a wavelength independent manner the relative power and phase in such distribution network. This naturally lead to a layout based on a slab coupler. Next, the optical length from each tap to the corresponding GC, $\Delta L$, can be either equalized or set to a controlled difference. The first approach ideally sets the light in all GCs with the same optical phase, and then one should resort to tuning individually each path from the slab coupler to the GCs. Once more this does not scale up well with $N_{GC}$, more when considering power consuming tuners such as thermo-optic phase shifters, which furthermore have a limited operating speed of a few tens of kHz (that is ultimate related to the system frame rate for a given number of points in the acquired scene). An architecture with path length increment $\Delta L \neq 0$ such as the one proposed in \cite{Van:11} scans in both H and V with the TL. However, bringing upfront the requirement to use state of the start devices, e.g. a TL with wavelength scan range, $\Delta\lambda_{TL}$ in the optical C-band of 35~nm, results in very long path length increments.

In summary, all the above points to a large number of elements / area OPA with a considerable number of optical paths and anticipated long lengths. 

\subsection{Technology}
Thus, among the most common monolithic photonic integration technologies, we selected silicon nitride that features the lowest propagation loss. Among the platforms available \cite{Munoz:17}, we employed the technology by Ligentec \cite{Munoz:19}. Once the technology is selected, there are several additional implications for the blocks used to construct the OPA. 

Firstly we considered the GC implementation in such platform. As reported in \cite{Kim:19,Raval:17,Poulton:16}, long and weak GCs are required, constructed with waveguides having a lateral etch. Our simulations, confirmed experimentally as we describe later in the paper, indicated that the GC divergence was in the range of 0.08\dg/nm. For $\Delta\lambda_{TL}$ that results into a FOV$_V$=2.8\dg, far from the target set above. The conclusion at this point leads to a switched architecture, in which OPAs having GCs with different period are used, as we will show later on. Each would cover a given part of the FOV in V, and we term them vertical blocks (VB).

Secondly, increasing the FOV in H is accomplished by placing the GCs close enough. The GC relative spacing needs to follow $d_{GC}=1.9\lambda$ to ultimately meet the 30\dg FOV$_H$ set at the start. The angular separation of the OPA diffraction orders in H is inversely proportional to $d_{GC}$. This is totally analog the the concept of free spectral range (FSR) in an arrayed waveguide grating (AWG), as described in \cite{Munoz:02:jlt}. Thus, one concern when aiming at increasing FOV$_H$ was the optical coupling of long closely spaced GCs running in parallel, that we investigated by means of test structures as we show later on.

\begin{figure*}[bt]
\centering
\includegraphics[width=1\linewidth]{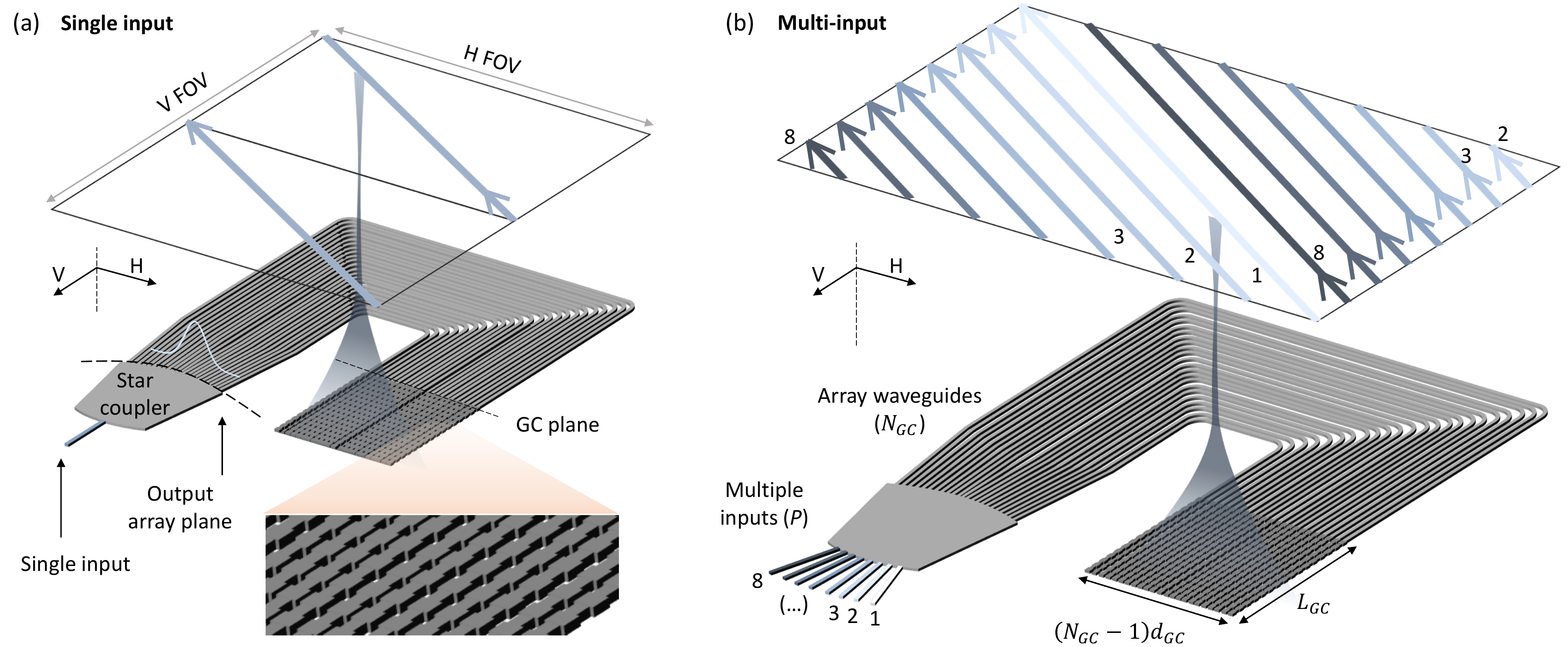}
\caption{(a) Single-input star coupler OPA, for which there is a single diagonal scanned field line that replicates for different orders in V. (b) Multi-input star coupler OPA, same device with multiple inputs, providing a $P$ times denser scanned field.
}
\label{fig:multi}
\end{figure*}

Thirdly, and as a consequence of the first remark above, stitching VBs by placing different GC design OPAs has footprint implications: the larger the footprint for each OPA is, the lower number of VBs in a given chip area will be. The OPA radiating elements block footprint is bounded to the GC length, $L_{GC}$, times the $(N_{GC}-1)d_{GC}$. However we cannot neglect the footprint of the distribution network, that is the array of waveguides from the slab coupler to the GC set. These have a relative length increment of $\Delta L$ as introduced above. In the simplest case, in which one of the diffraction orders in H is shifted the full FOV$_H$ by sweeping the wavelength of the TL an amount of $\Delta\lambda$, the length increment needs to satisfy $\Delta L = \lambda^2 / (n_g \Delta\lambda)$, where $n_g$ is the waveguide group index. As shown in Fig.~\ref{fig:multi}(a), a single line would be drawn by the steerer through a complete H FOV and the VB FOV. It is possible to have more than one diffraction order scanning in H within the same VB FOV. In that case, $\Delta L = \lambda^2 / (n_g ( \Delta\lambda / N_{cycles}) )$, where $N_{cycles}$ is the number of diffraction orders drawing lines within the VB FOV. For the single input SC-OPA being now described, the number of lines drawn is equal to cycles $N_{\text{lines}}=N_{\text{cycles}}$. As example, with $N_{lines}=2$, and the laser swept the same $\Delta\lambda$ than before, a first diffraction order would draw a line covering the full H FOV and half of the VB FOV, and a second diffraction order would then come into the block to cover the second half of the VB FOV, with a new line spanning the complete H FOV, as depicted in Fig.~\ref{fig:multi}(b). In conclusion, populating the VB with more lines, so as to increase the resolution in V, results into an increased footprint (large $\Delta L$). With $\Delta\lambda$=35~nm, 0.1\dg V beam width and 2.8\dg VB FOV densely populated results into $N_{lines}=28=2.8$\dg$/0.1$\dg, which leads to $\Delta L$=915.24~\um. In connection with the H direction, the beam width requirement is 0.1\dg as well. This is related to $(N_{GC}-1)d_{GC}=1$~mm and FOV H is inversely proportional to $d_{GC}$ as mentioned above. For the FOV H of 30\dg, $d_{GC}=1.9\lambda = 2.94$~\um. Then $N_{GC} \simeq 340$, so the length difference between the shortest and longest waveguide would be approximately 31~cm. This is just for a single OPA covering a VB FOV of just 2.8\dg. Placing some of these in parallel within the typical fabrication reticle of lithography steppers, it's extremely challenging if not impossible. 

\subsection{Architecture}
Faced to this roadblock, in fourth place we explored an additional degree of freedom within the SC-OPA device. All the above refers to a device using a single input to the SC. The SC far-field phase for light coming from different SC inputs can differ by a constant amount. It can be designed and controlled so as to span a range of $2\pi$ for $P$ inputs. Consider then a SC-OPA with $P$ inputs, whose positions are engineered in such a way that the SC far-field phase at a given wavelength, for two adjacent inputs, differs $2\pi/P$. For a fixed (not scanning) wavelength, the spot produced in the scene, will be shifted in H an amount $FOV_H / P$ when switching between consecutive inputs, while staying in the same V position within the VB. This is depicted in Fig.~\ref{fig:multi}(b). When sweeping, the trajectories will not overlap, but interlace among them as shown as well in the same picture. However, the most remarkable outcome is that for the multi-input SC-OPA, the length increment can be re-written as:

\begin{equation}
\label{eq:DL}
\Delta L = \frac{\lambda^2}{n_g\frac{\Delta\lambda}{N_{\text{cycles}}}}.
\end{equation}
with $N_{\text{cycles}}=N_{\text{lines}}/P$. As numerical example to compare with the previously presented, consider a multi-input SC-OPA with $P$=14 to draw the same 28 lines ($N_{\text{cycles}}=2$ diffraction orders). Then the required $\Delta L$ for the same TL tuning range of 35~nm would be 65.37~\um. For each input, 2 lines are drawn in the FOV. The H position shift due the the SC relative phase for the inputs, ensures the next 2 lines for the next input do not overlap in the FOV. When considering the ultimate target of H beam width and FOV, the length difference between the shortest and longest waveguide would be now reduced by the same factor $P$, that is 2.2~cm, comparatively a less daunting challenge.

Last but not least, the slab coupler (SC) element feature power distribution with ideally equalized phase at the start of all the paths to the GCs. For silicon nitride, as compared to very low index contrast silica on silicon technology, the arrangement of the SC outputs is done as in \cite{Zou:17}.

In summary, the proposed switched architecture, with interlaced scene scan, features smaller footprint for the OPA, as compared with a single input device. Details on the footprint advantage, including the switching matrices, are elaborated in section~\ref{sec:test}.

\section{\label{sec:dsgnfab}Design and fabrication}

As indicated in the previous section, we resorted to the silicon nitride AN800 platform by Ligentec, whose details can be found in \cite{Munoz:19}. In short, a 0.8~\um thick silicon nitride embedded in silica is the guiding layer. Besides this basic function, a heater module and a deep trench module were used as well. In our designs, the basic waveguide is of $W_{wg}=0.8$~\um width.

\subsection{Grating couplers}
With this waveguide cross-section, we firstly addressed the design of the long GCs. We resorted to a single etch step for the silicon nitride layer, thus the gratings were patterned as lateral corrugations of a regular waveguide. This way, the GC period is comprised of two sections of different widths,  $W_{a,b}=W_{wg}\pm \Delta w_{GC}$. In order to cover several VBs as outlined in the previous section, GCs with different period length, $\Lambda_{GC}$ were designed as well. A design of experiments (DoE) from previous building block (BB) development runs comprised gratings of periods of $[842.39,872.23,904.32,938.85,976.07]$~nm for target angles from -5.0\dg to +5.0\dg in steps of 2.5\dg (with 0\dg being normal to the chip plane). The lateral etch was in $[15,30,45,60]$~nm. Uniform and apodized versions of these GCs were explored. The apodization profile (etch-depth variation along the GC length) is numerically determined for the near field to have Gaussian shape with minimum truncation that guarantee a given Main to Secondary Lobe Ratio (MSLR) in the vertical direction, while maximizing energy transfer from the waveguide to free-space. 

\subsection{Slab coupler}
Secondly, the design of the slab coupler was faced following the findings in \cite{Zou:17}. The authors therein demonstrated improved AWG imaging performance by placing the waveguides at the output of the SC at a constant distance over the tangent line to the SC circle, rather than at a constant angle over that circle as traditionally done with low index (silica on silicon) integration technologies. For our multi-input SC, we determined the location of the input waveguides by iterative simulation of the expected far-field horizontal pointing angle as follows: 1) we set the focal length (SC length) and output waveguide (OW) separation as described in \cite{Zou:17} that minimize the non-linear phase distortion along the $N_{GC}$ outputs when lateral inputs on the SC are employed; 2) for each input waveguide, its proper position along the input arc is determined simulating the amplitude and phase profiles projected over the output waveguides, and from these values the expected far-field beam pointing angle. In a short number of fast iterations the positions of the P input waveguides are determined to point towards the desired horizontal FOV angles. In our casem, these are P directions separated FOV$_H$/P. Steps 1 and 2 can be repeated modifying the focal Length, the number of OWs and OW separation, in order to adjust key features in the far-field, such as the beam width and the Main to Secondary Lobe Ratio (MSLR). Both are related with the shape and the overall SC far-field truncation to be allowed \cite{Munoz:02:jlt}.

\subsection{Switch unit cell}
In third place we designed the unit switch cell, based on a balanced Mach-Zehnder Inteferometer (MZI). This builds upon a previously designed 2x2 MMI available in the process design kit (PDK) for the AN800 Ligentec technology. Two thermo-optic heaters were used, one per MZI arm. Anticipating to the well-know thermal cross-talk effect for neighboring heaters, we did also use the deep trench module. Our switch cell layout has three of these trenches, one placed in the symmetry axis of the MZI, between the two arms, and two other at the opposite side of each arm. 

\subsection{Optical phased array}
With regards to the OPA design combining the BBs above, we have previously reported as proof-of-concept (PoC) \cite{Munoz:21} a single VB multi-input SC-OPA with no switching on-chip, some of its characteristics reproduced hereby for completeness and ease of comparison with the new design reported in the present paper. The previous design was for a H$\times$V FOV of 15\dg$\times$2.8\dg, using a TL with  $\Delta\lambda_{TL}=35$~nm, and a beam size is 0.36\dg$\times$0.18\dg . The GC separation was $d_{\text{GC}}=6$ \um. together with the $N_{GC}=39$ GC emitters, and apodized GCs of $L_{\text{GC}}=1$ mm. These result into an OPA aperture of (H$\times$V) 234~\um~x~1~mm. Last but not least, the design features 2 horizontal lines within the VB FOV per device input, that is $N_{\text{lines}}=16$ and $P=8$ and therefore by using Eq. \ref{eq:DL}, $\Delta L = 65.4$~\um. We labeled this design as SC0.

In this paper we additionally report on a new design iteration, labeled SC1, with some of the features changed, and fabricated in a different run than SC0. The H emission area was doubled by increasing the number of emitters to 78, while keeping $d_{\text{GC}}=6$ \um, with the purpose of reducing the beam width in H. We kept the same target H$\times$V FOV of 15\dg$\times$2.8\dg. The V resolution was set to 0.23\dg, which then requires $N_{\text{lines}}=12$ and $P=12$, so $\Delta L = 32.7$~\um. However, we just laid out 8 of the 12 inputs, aiming at a more compact switch matrix stage (with number of lines a power of 2). Consequently, and as shown later, parts of the FOV are not covered, but there is not a fundamental limitation as we demonstrated and show hereby for the SC0 design. In addition, one aspect was to keep the longest path difference equal in both the SC0 and SC1 designs, which is met by halving $\Delta L$ while doubling $N_{GC}$. The motivation behind was to investigate the addition of more radiators, in terms of the phase distortions added, but with the same path length. Phase errors stem from non-uniform fabrication of the waveguides, mainly due to variations in waveguide material composition and waveguide thickness (deposition effects) and waveguide sidewall roughness (lithography and etching effects). Design improvements such as waveguide up/down taper in the straight sections of the waveguides are common place, such as in \cite{Dostart2020}. Large scale OPAs in silicon nitride are also reported with 4×4~\mms \cite{Poulton:17} and 4x3~\mms \cite{Li:21} showing the current technology limits for phase coherence on waveguide arrays. Four SC1 OPA designs, each with GCs having a different period, were include within the same chip, together with the switching matrix stages to select individual inputs among the ones for each OPA, and to switch among VBs. The layout is shown in Fig.~\ref{fig:sc1_chip}(a) later on.

In conclusion, two SC-OPA designs are considered, termed SC0 and SC1. They differ in the number of radiating elements (two times for SC1), that is the OPA aperture, and consequently the target beam width in the horizontal FOV. Another key difference is the number of horizontal lines per VB, which for SC0 is 2, while for SC1 is 1, implying $\Delta L$ is half for the latter, so the length difference between the shortest and longest path to the GCs is the same for both designs.

\section{\label{sec:tech} Technology and building block characterization}

\begin{table}
\caption{Technology figures of merit}
\begin{tabular}{|c|c|c|c|}
\hline
\multicolumn{ 1}{|c|}{\textbf{Building block}} & \textbf{Symbol} & \textbf{Unit} & \textbf{Value} \\ \cline{ 1- 4}
\multicolumn{ 1}{|c|}{I/O taper} & IO$_{TE}$ & dB & 0.87-0.93 \\ \cline{ 2- 4}
\multicolumn{ 1}{|c|}{} & IO$_{TM}$ & dB & add 3 to TE \\ \cline{ 1- 4}
\multicolumn{ 1}{|c|}{Straight waveguide} & $\alpha_{TE,sw}$ & dB/cm & 0.183 \\ \cline{ 2- 4}
\multicolumn{ 1}{|c|}{} & $n_{eff,TE}$ & - & 1.717-1.719 \\ \cline{ 2- 4}
\multicolumn{ 1}{|c|}{} & $\alpha_{TM}$ & dB/cm & 0.165 \\ \cline{ 2- 4}
\multicolumn{ 1}{|c|}{} & $n_{eff,TM}$ & - & n.a \\ \cline{ 2- 4}
\multicolumn{ 1}{|c|}{} & $n_{g,TE}$ & - & 2.08-2.12 \\ \cline{ 2- 4}
\multicolumn{ 1}{|c|}{} & $n_{g,TM}$ & - & 2.08-2.12 \\ \cline{ 1- 4}
\multicolumn{ 1}{|c|}{Bent waveguide} & $\alpha_{TE,bw}$ & dB/cm & 0.876 \\ \cline{ 2- 4}
\multicolumn{ 1}{|c|}{} & $\Delta\beta$ & 1/µm & 1.1e-3 \\ \cline{ 2- 4}
\multicolumn{ 1}{|c|}{} & $\kappa_P$ & 1/µm & 3.75e-4 \\ \cline{ 1- 4}
\multicolumn{ 1}{|c|}{Spiral waveguide} & $\alpha_{TE,sp}$ & dB/cm & 0.52 \\ \cline{ 2- 4}
\multicolumn{ 1}{|c|}{} & $\alpha_{TM,sp}$ & dB/cm & 0.40 \\ \cline{ 1- 4}
\multicolumn{ 1}{|c|}{Optical coupler} & K$_{TE,50:50}$ & - & 48.3:51.7 \\ \cline{ 2- 4}
\multicolumn{ 1}{|c|}{} & K$_{TM,50:50}$ & - & 47.2:52.8 \\ \cline{ 2- 4}
\multicolumn{ 1}{|c|}{} & K$_{TE,85:15}$ & - & 82.1:17.9 \\ \cline{ 1- 4}
\multicolumn{ 1}{|c|}{Grating coupler} & Type & - & Uniform \\ \cline{ 2- 4}
\multicolumn{ 1}{|c|}{} & Angle & deg & 0 \\ \cline{ 2- 4}
\multicolumn{ 1}{|c|}{} & $\alpha_{GC-TE}$ & dB/mm & -2.15 \\ \cline{ 2- 4}
\multicolumn{ 1}{|c|}{} & $\alpha_{GC-TM}$ & dB/mm & -7.04 \\ \cline{ 1- 4}
\multicolumn{ 1}{|c|}{Thermal tuner} & I$_p$ & mA & (T) 41.91 // 
(NT) 44.55   \\ \cline{ 2- 4}
\multicolumn{ 1}{|c|}{} & t$_r$ & ms & (NT) 38 \\ \cline{ 2- 4}
\multicolumn{ 1}{|c|}{} & BW & kHz & (NT) 9.2 \\ \cline{ 2- 4}
 & X$_{NT}$ & \% & 12\% \\ \cline{ 2- 4}
 & X$_T$ & \% & 2.5\% \\ \hline
\end{tabular}
\label{tab:tech}
\end{table}
Firstly we addressed on the technology performance. The results are given in Table~\ref{tab:tech}. The results are presented for TE and TM, despite our device design and experimental validation in section~\ref{sec:test} is only for TE. However, some of the unwanted features observed in the performance are to be attributed to TE-TM interaction. The input and output (I/O) tapers insertion loss are characterized by coupling light to a set of straight waveguides equipped with such tapers, relative to the setup with no chip between microscope objectives. The polarization filters in the in/out coupling stages of the setup were set to TE or  as needed. For sets of waveguides, we found the insertion loss in the range of 0.87 to 0.93~dB for TE, and extra 3~dB for TM.

The features for the straight waveguides (width 0.8~\um, height 0.8~\um) were derived from multiple test structures. As core test structure (TS) for both the basic properties of waveguides and multi-mode interference (MMI) couplers, we used an unbalanced MZI with arms having sufficient length difference to resolve the optical losses. The design follows our previously reported approach in \cite{Bru:19}, where the technique based on optical frequency domain interferometry (OFDI) measurements permits to obtain the waveguide propagation losses, and the results are given within the same Table~\ref{tab:tech}. The group index was derived from the spectral trace of unbalanced MZIs (where the length increments were only on straight waveguide pieces). Finally the effective index (for TE) from the center wavelength of the transmission response from a test GC emitting at 0\dg, that was recorded on an OSA. 
The bent waveguide (for the same cross-section and bend radius of 150~\um) features were derived from a ring-resonator TS, constructed with 50:50 MMIs. The OFDI response (set of pulses with decreasing intensity) was fitted, the loss for the straight parts and MMI splitting ratio deducted, and thus the one for the bends derived. The coupling between the TE and TM mode, following the model in \cite{Mori:06}, was also extracted from the OFDI response, where after some re-circulations TM pulses arose in between TE pulses locations. Spiral waveguides, of length 5~cm, and waveguide spacing of 20~\um, with a 33\% of bends of the same radius, were used as to cross-check the derived magnitudes for straights and bends from other structures.

Unbalanced MZIs were used as TS to asses on the thermal tuner efficiency. This TS had thermal tuners immediately on top of the arm waveguides. Then, for one of the arms, additional tuners were laid out at multiple distances of 10~\um. On the other hand, a similar layout was used for the other arm, but one out of two heaters well replaced by deep trenches, that is the arm heater was separated by 30~\um of the next heater, with a deep trench between them. The spectral shift of the MZI was measured for a current range of 0-60~mA, for every heater of the described above. More specifically, for the lateral displaced heater at 30~\um without trench the relative spectral shift was 12\% respect to the direct heater. In the same geometrical, but with a trench at 15~\um (middle point) we measured a relative displacement of 2.5\%. The MZI switches use thermal tuners, with current to shift $\pi$ of 41.91~mA for those equipped heat isolation trenches, as detailed in Table~\ref{tab:tech}. Electrical test structures having same pads dimensions, and incremental track lengths (of same cross-section than the actual heaters) were measured. From these, the heater resistance was estimated to be 34~Ohm, and thus the power consumption for a $\pi$ phase shift was inferred to be $\approx$60~mW.

\begin{figure}
\centering
\includegraphics[width=1\linewidth]{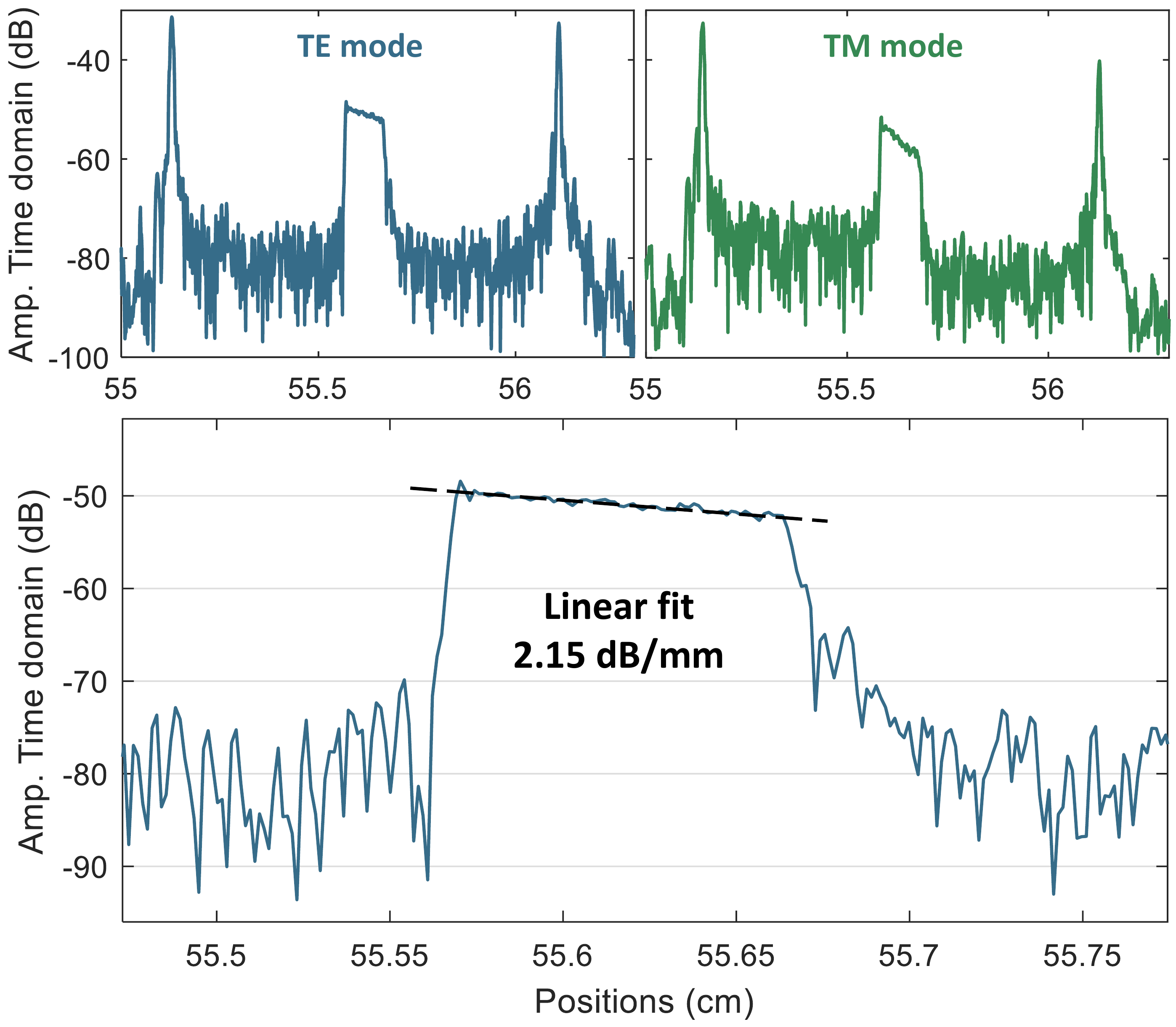}
\caption{(top panel) TE and TM mode OFDI reflectometry traces for grating couplers. (bottom panel) zoom-in view in the grating location for the TE mode.
}
\label{fig:ofdr_gc}
\end{figure}

With the OFDI setup, we did also retrieved the reflectometric time domain response and obtained the decay rate of energy for the GCs, as shown in Fig.~\ref{fig:ofdr_gc}. The results were -2.15 and -7.04~dB/mm for TE and TM respectively, for the GC having design parameters $\Lambda=904$~nm and lateral etch depth of 60~nm.

Finally, some TSs were used to explore the coupling strength of GCs running in parallel. These comprised pairs of GCs placed at distances from 1.2 to 3.0~\um. We found that the cross-over length of such directional coupler-like structure, for the GCs at 3.0~\um distance was above 88~cm, that is negligible coupling for GCs of 1~mm at the designed $d_{GC}=6$~\um.

\section{\label{sec:test}OPA chip characterization}
\subsection{Approach}

\begin{figure}
\centering
\includegraphics[width=1\linewidth]{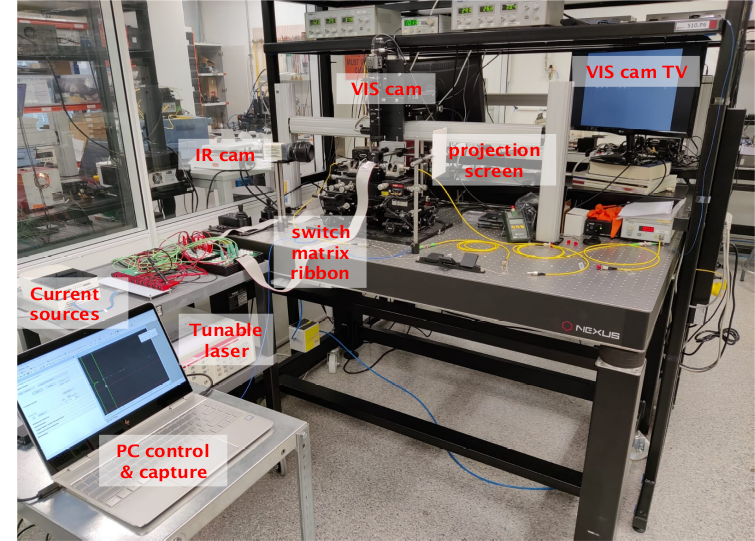}
\caption{OPA chip characterization setup}
\label{fig:setup}
\end{figure}

Our different OPA proposals were experimentally tested in the laboratory after refining designs, fabrication processes and several BBs. As described in the previous sections, two were the main devices so as to prove the interlaced scanning patterns: designs SC0 and SC1. For this, we assemble the experimental setup shown in Fig.~\ref{fig:setup}. 

We performed edge coupling by using microscope objectives as mechanism for feeding light into the PIC. In order to control the input polarization, polarization controllers are employed previous to the microscope objectives, filtering TM mode and letting TE mode be coupled. In the case of design SC0, as the scan is performed with a purely passive method (we directly access the SC input ports), the PIC is deposited on top of a copper chuck which holds the die through a vacuum channel.
For the SC1 design, the scanning pattern is accomplished by controlling the switching matrix to select the operating VB and the corresponding input to the SC. Thus, this PIC was assembled in an aluminium PCB, where wire-bond connectors link the DC pads contained in the PIC with the pads contained in the PCB.
Moreover, as heat dissipation is crucial in this case, the PCB is fixed on top of an aluminium chuck and a heat radiator. To control the temperature changes happening in real time in the PIC, a ceramic thermistor is also included and routed to the PCB terminals. The thermal management is performed by means of a thermal electric controller (TEC), which maintains a constant temperature through the characterization process using a Peltier cell. In other hand, a proper arrangement must be done to capture the light beam generated by the OPA. We employed a infra-red (IR) CCD 2D array placed 14~cm over the PIC
for the acquisition of the beam profiles with enough definition ($\approx$10 pixels along the main beam). This distance is just slightly larger than the far-field distance for the apertures considered, but via simulations we confirmed this have no substantial effects. Next, in a second arrangement, and in order to capture the complete FOV, we resorted to a mirror and projection screen. For this, a 90\dg gold protected mirror (96 \% reflectance) is allocated above the PIC (4cm) and a viewing screen is mounted and adjusted in height to collect the beam reflected. At the other end of the setup, the InGaAs CCD array  (Hamamatsu C12741-03, 14 bit resolution, 42 dB dynamic range),
is mounted with an IR objective and prepared to capture the viewing screen. Processing includes dark frame subtraction to eliminate laboratory ambient light, and gamma correction.
After the adjustment and characterization of this setup, the area covered with the InGaAs camera was 22.5\dg H x 17\dg V. The difference between the horizontal and vertical axis is due to the detector array of the InGaAs camera (640x512 pixels). This range is enough to properly capture the scanning patterns created by designs SC0 and SC1. 

In the case of design SC1, the switching matrix must be electrically powered up. For this, two FFSD 40-pin flat cables are connected to both sides of the PCB and to an interconnection 40 points electrical box, where several DC lines can be easily switched to activate different heaters from the switching matrix. Finally, the electrical interconnection box is connected through an interface to a couple of 10-channel, 300~mA current sources provided by LuzWaveLabs. This current sources are especially designed for PIC testing, able to provided 16~bits of resolution, high stability and excellent noise performance. For their control, several Python and MATLAB scripts were generated.

\begin{figure}
\centering
\includegraphics[width=0.45\textwidth]{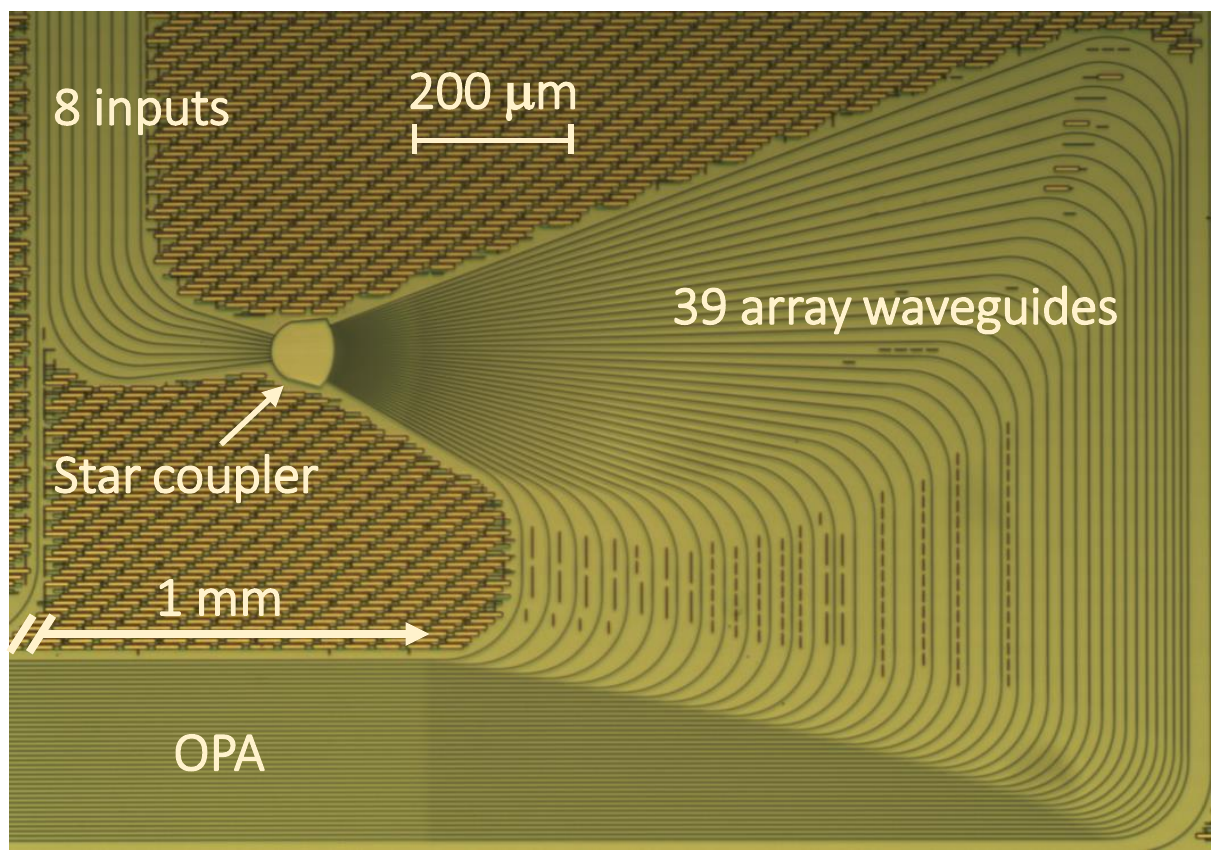}
\caption{
Multi-input star coupler fabricated device microscope picture.
}
\label{fig:sc0_chip}
\end{figure}

The primary objective then is to capture the scanning pattern of the OPAs using the experimental setup previously described. In the case of SC0, it contains a single VB where the multi-input SC has 8 inputs which were accessed individually from the chip facet. For the SC1 design, the 4~VBs can be selected by a 2x4 switching matrix followed by a 2x8 for each of the VBs. Thus, the first task to generate the OPA scanning pattern is to calibrate the different switching matrices. It is worth mentioning that the tuning of each phase shifter contained in the MZIs (just one of the two arms was tuned) was done by performing a current sweep of 0-60~mA, range in which a $\pi$ phase shift is seen. This information was obtained by previous BB individual characterization. For each VB light is edge coupled to the upper 2x2~MZI of its corresponding switching matrix. Note that in both designs, SC0 and SC1, the not needed MZI inputs at all stages of the switching matrix are routed independently to the die facets on purpose, so they can be employed in the switching matrix calibration phase. In short, once light is applied to one of the sparse 2x2 MZI input we can observed using the camera both radiated spots, and tuning the phase shifter light can be redirected to just one of the spots (i.e. one inputs of the SC). This step is repeated for the each sparse 2x2 MZI input. This task has been developed manually. Nevertheless, the process can be easily adapted to be performed through image recognition and automated edge coupling.

After the experimental setup preparation and the switching matrix calibration, the OPA interlaced scanning patterns can be properly tested. In the case of design SC0, light is edge coupled to the desired VB through the outer 2x2~MZI from the switching matrix. Similarly, in the case of design SC1, light is edge coupled to the VB selector, located in the left lower part of the die (see Fig. \ref{fig:sc1_chip}).
Next, the 90º gold mirror is positioned and adjusted on top of the die. After this, the OPA beam is visible in the viewing screen allowing a fine tuning and maximization of the edge coupling by using a tuneable laser and a manual polarization controller. Once TEC controller is set to a temperature of 25\dg and the ambient light compensation frames are captured each stage of the switching matrix is properly set to drive light to the first input of the SC, generating a light beam in the viewing screen and capturing the video frames as wavelength is swept using the tuneable laser. The process is repeated adjusting the switching matrix to drive each SC input and restarting the wavelength swept.
Once the VB has been completed, the electrical connections from the interconnection box need to be arranged to feed the next VB. In the case of design SC1, the VB selector is properly modified. Similarly, the switching matrix of the VB is prepared to drive light to the first input of the SC and the measurement is repeated as previously described. After the recording of the interlaced scan patterns generated by the OPAs, the videos are processed and results are extracted using our own MATLAB routines. 

\begin{figure*}[t]
\centering
\includegraphics[width=0.24\textwidth]{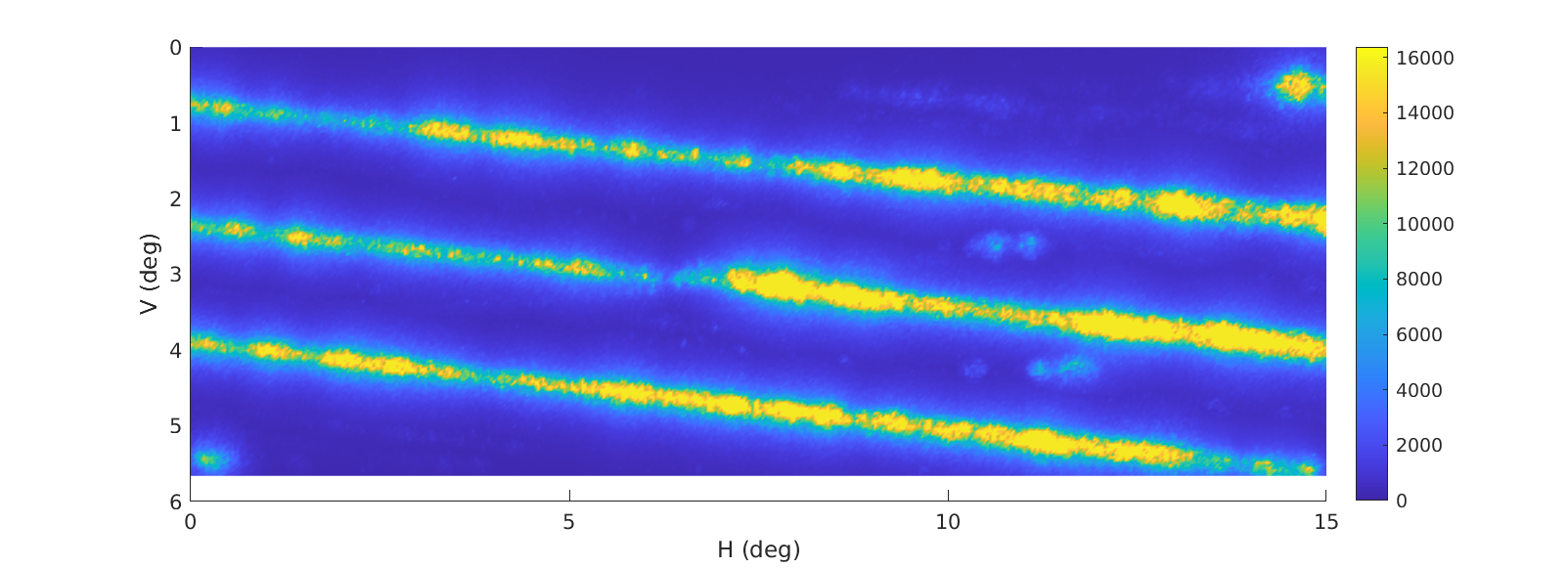}
\includegraphics[width=0.24\textwidth]{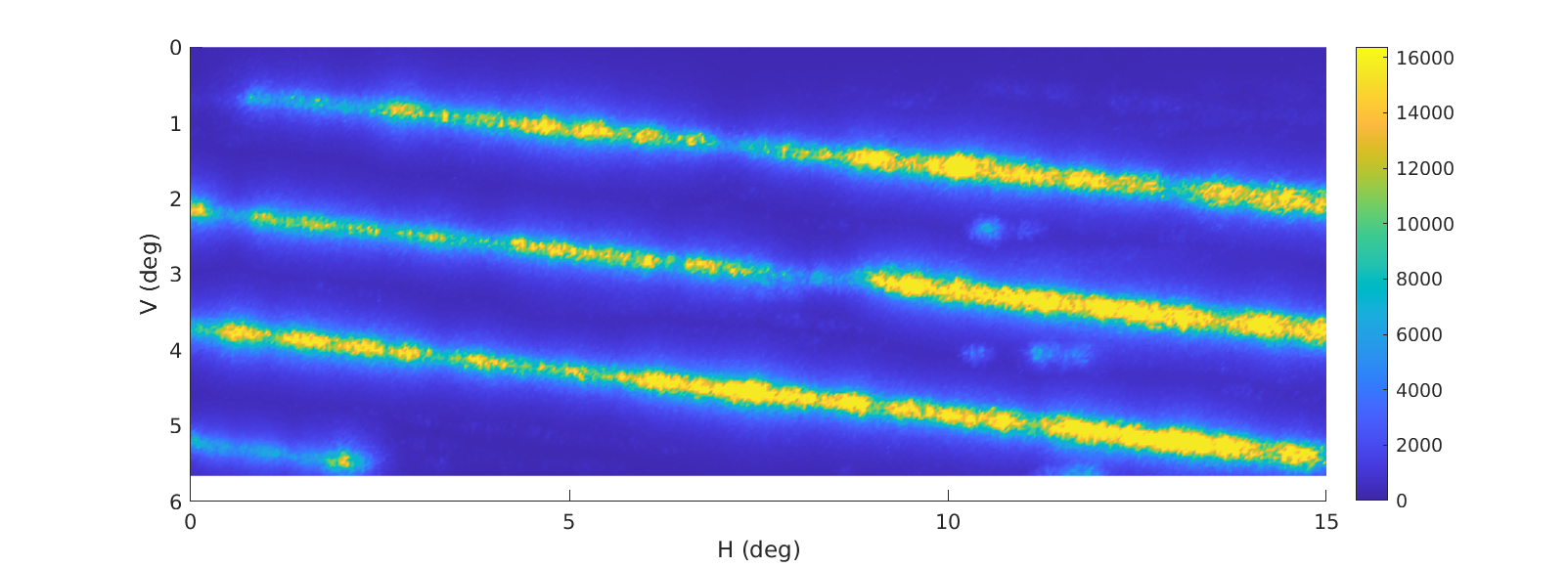}
\includegraphics[width=0.24\textwidth]{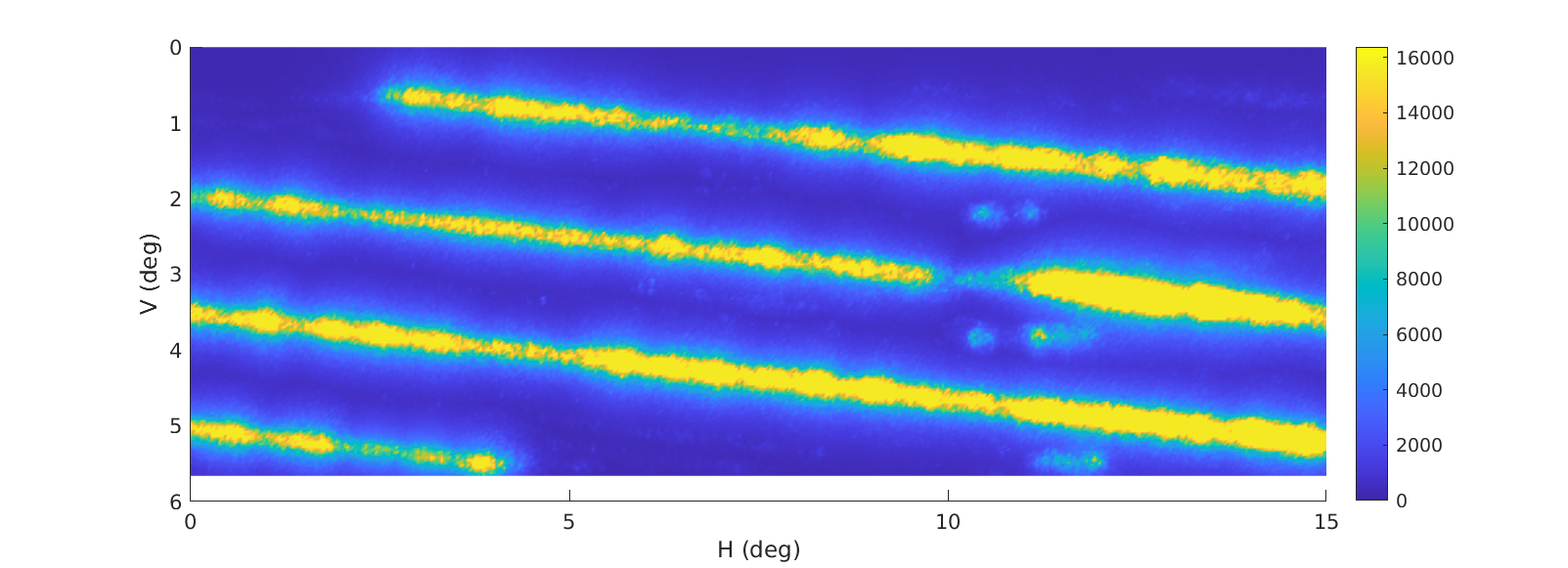}
\includegraphics[width=0.24\textwidth]{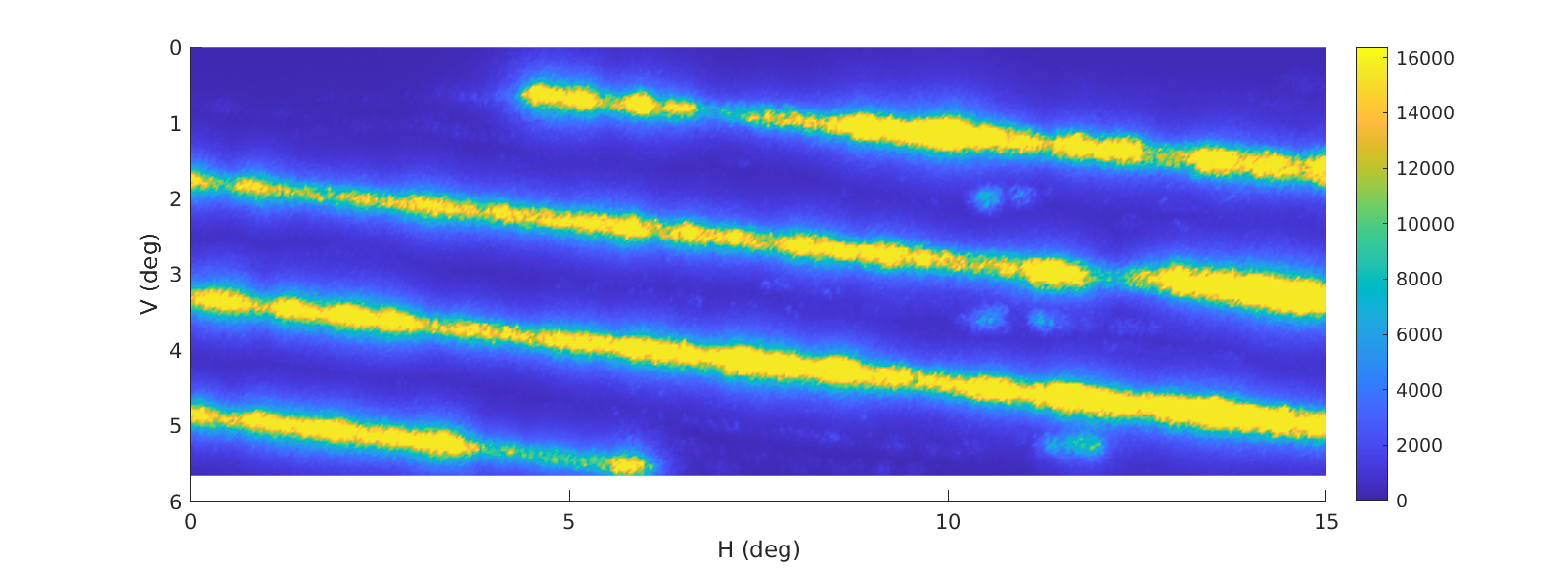}

\includegraphics[width=0.24\textwidth]{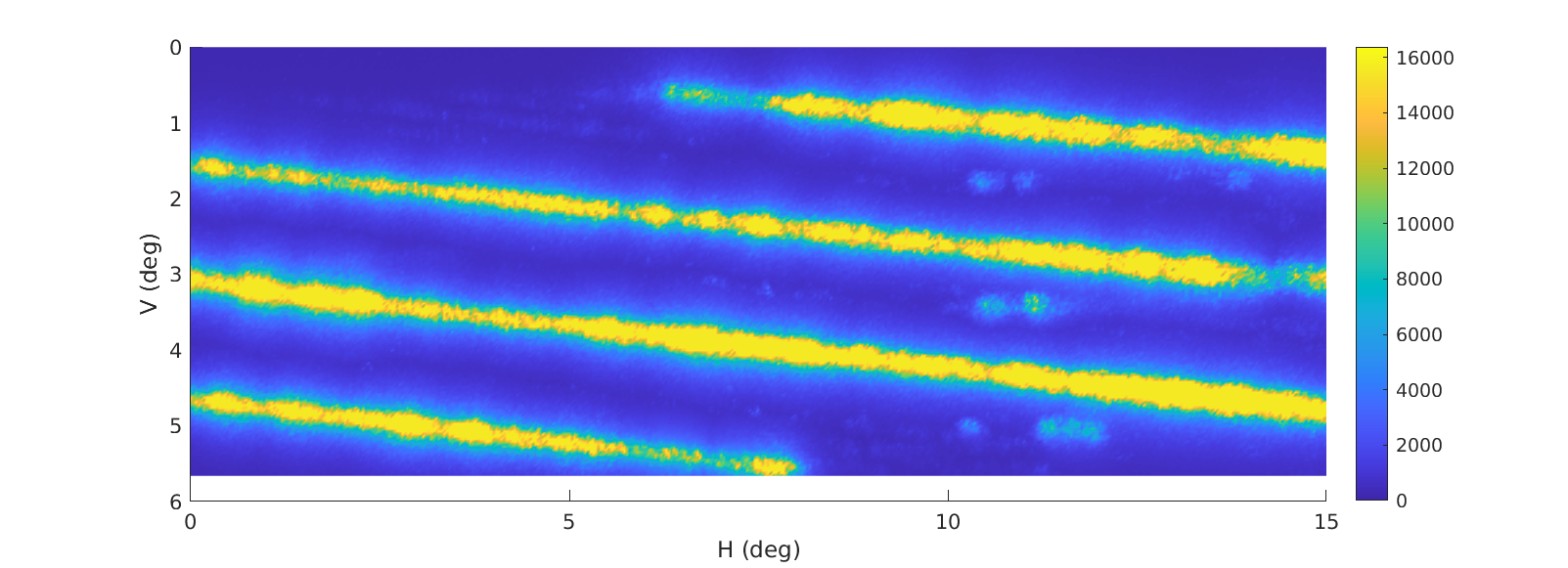}
\includegraphics[width=0.24\textwidth]{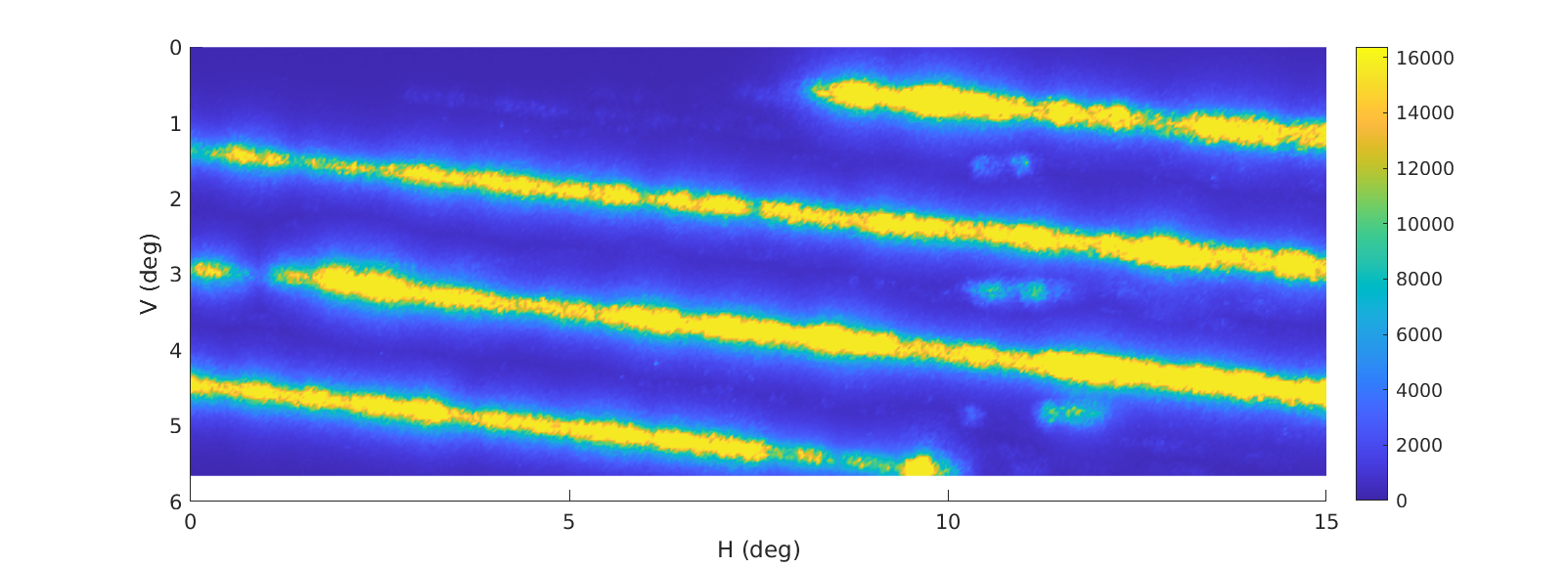}
\includegraphics[width=0.24\textwidth]{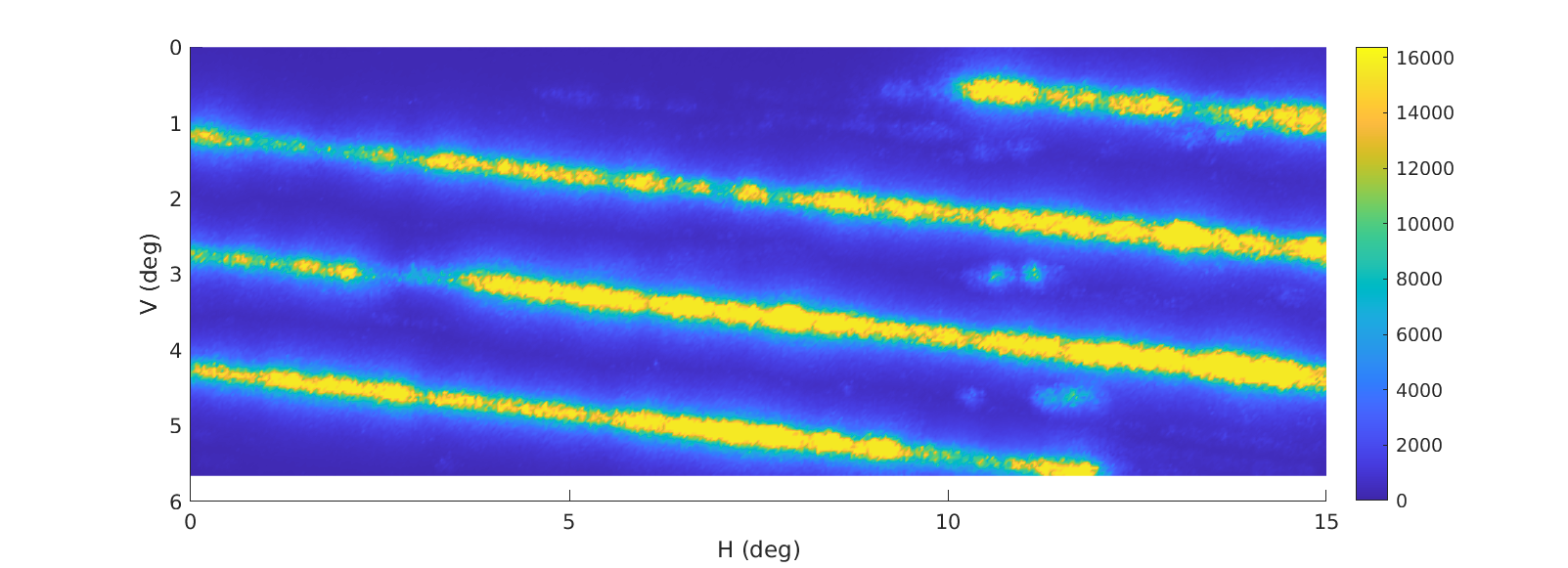}
\includegraphics[width=0.24\textwidth]{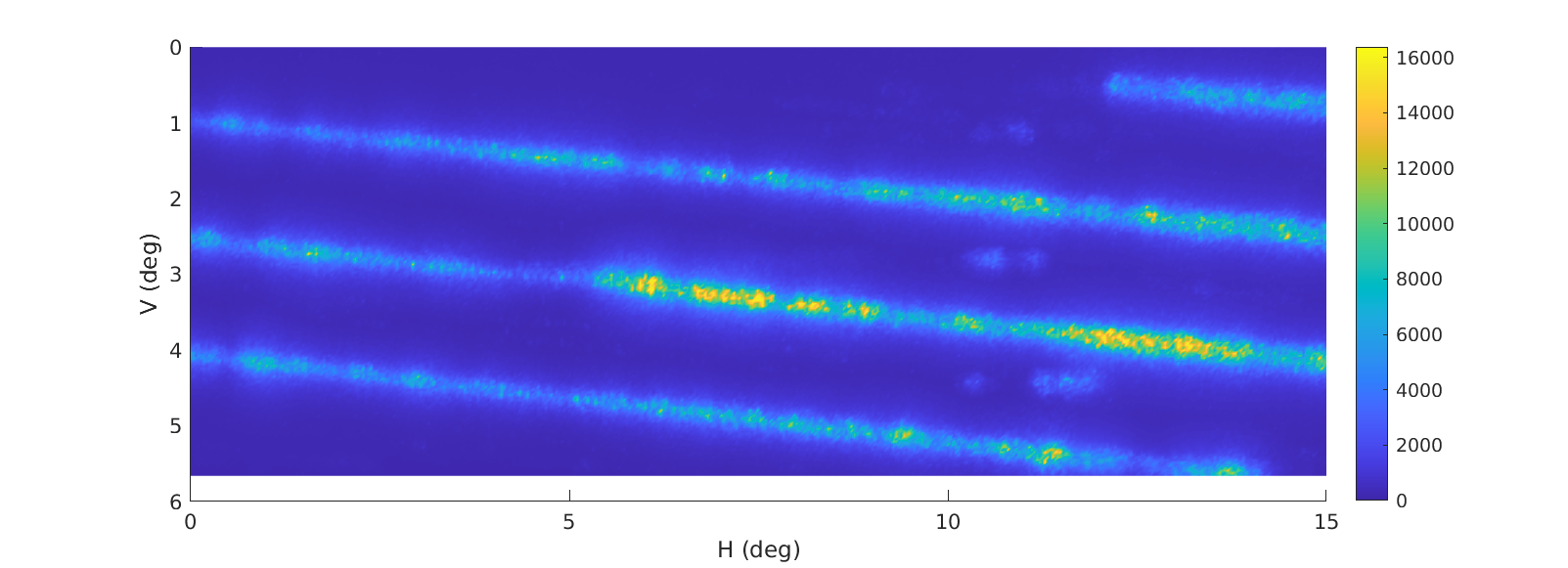}

\includegraphics[width=1.0\textwidth]{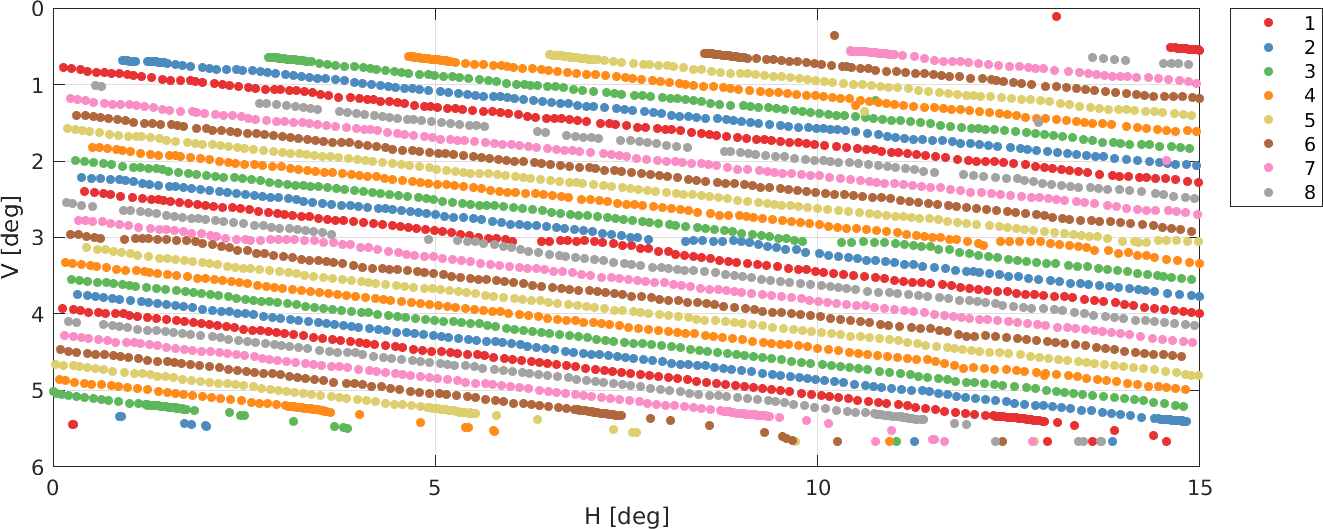}

\caption{
SC0 chip device results.
(top 8 small plots) Video frame-integrated plots of the scan field for the 8 inputs, showing the oblique line pattern for each case.
(bottom large plot) Measured peak intensity positions, showing the multi-input interlacing (inputs 1--8, in pink--grey color) that fills the scan field.
}
\label{fig:exp}
\end{figure*}

\subsection{Experimental results}

We firstly expand the results for our SC0 chip, previously reported in \cite{Munoz:21}. A picture of the SC0 chips is given in Fig.~\ref{fig:sc0_chip}. Light from the TL is horizontally coupled to each of the 8 inputs, which are routed to the chip facet. The TL scan range was in this case 50~nm (1525 to 1575 nm), larger than the 35~nm to cover the FOV. The overall capture strategy entails video recordings, and allows each video frame to be linked to a given wavelength of the sweep. The projection of the OPA spots on the screen was then recorded, and each video was off-line processed. After capturing videos for each input, we first integrate by collapsing all the frames into one image by holding the maxima. This gives a clear representation of the beam trace as it is represented in Fig. \ref{fig:exp}(a). At the beginning of the scan, the beam is placed in the upper left corner. As wavelength is swept, as it is observed the beam traces the expected diagonal path composed by the faster movement in H due to linear incremental phase changing in the array, combined with V one because of the GC dispersion. The next OPA diffraction order lobe appears in the field when 15\dg in H are steered, repeats the travel twice, and ends at the bottom. Same dynamics are observed for every input, but progressively displaced a fraction in the V dimension. This offset corresponds exactly to a $2\pi/8$ phase jump, so that all the input traces in the same order interlace homogeneously the space between two orders of the same input, as expected, to fill the entire scan field. In fact, there are observed two groups of 8 lines for $\Delta\lambda=35$ nm, covering the 2.8\dg intended for the vertical block. This result can be better observed in Fig. \ref{fig:exp}(b). There is observed some no ideal behaviour though. Along their trajectories, which are not exactly linear, beam widths vary and significant intensity fading is observed. 

\begin{figure}
\centering
\includegraphics[width=0.45\textwidth]{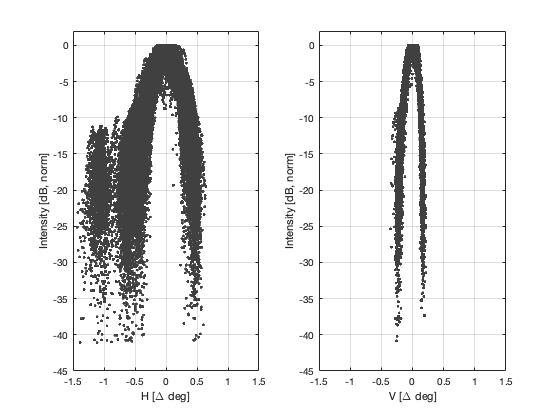}
\includegraphics[width=0.45\textwidth]{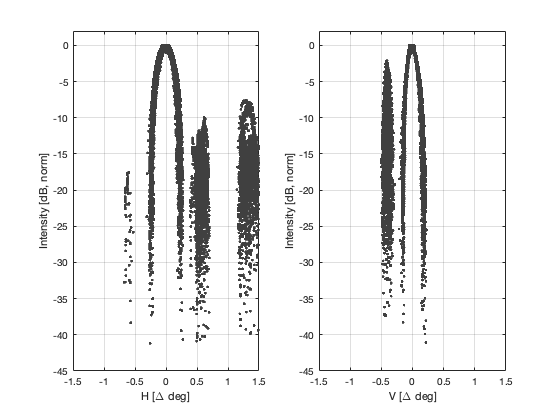}
\caption{Beam frame H (left) and V (right) slices for the SC0 (top) and SC1 (bottom) chips.
}
\label{fig:beams}
\end{figure}

Individual frame captured beams are fitted to a 2D Gaussian model, so the H and V widths are retrieved by computing its full width at half maximum (FWHM). We observed beam quality was degraded when capturing from the projection screen, as compared with direct beams captured with the IR cam on top of the chip. The beams from the projection screen exhibited glitches and peaks, we attributed to stray photons both from the screen and other light sources in the lab. These degrading artifacts were not observed for the IR cam on top of the chip. Consequently, we acquired beams following the same TL scan and video record procedure directly seen from the chip in the IR cam, at the cost of a reduced FOV. In Fig.~\ref{fig:beams}(a), all beam frame slices are represented together for H and V, at left and right, respectively. The beam width is estimated from the Gaussian fittings (excluding frames with beams having peak intensity below 20\% of the IR cam dynamic range) and matches with the design, 0.36\dg$\times$0.175\dg. The average beam wander is below one beam width though. For H, we observe the presence of a region at -1\dg from the beam centers where artifacts in between -10 and -30 dB from the peak maximum appear. By observing Fig.~\ref{fig:exp}(a) at 1\dg from the beam trail, the presence of the side artifact is present in a small part of the FOV. This might be attributed to TE-TM mode interaction in the waveguides, similar to what \cite{Kleijn:12} reports for AWGs. The peak intensity fluctuations can also be observed in Fig.~\ref{fig:exp}(a) for H and V. These phenomena can be attributed to coherent interference with chip intra-reflections: as it is well-known, diffraction in GCs have emission downwards that, when reflecting on the different die layer interfaces, interferes coherently with upwards emission, forming Fabry-Perot cavities. The buried oxide interface with silicon substrate produces a slow envelope, and the Si-air implies a larger cavity and thus, a faster variation. The observed variation periods are corroborated by simulation. This problem might be fixed by including anti-reflection coating, or a more complex dual layer unidirectional GC \cite{Raval:17}, which removes most of GC downward emission.

\begin{figure}
\centering
\includegraphics[width=1\linewidth]{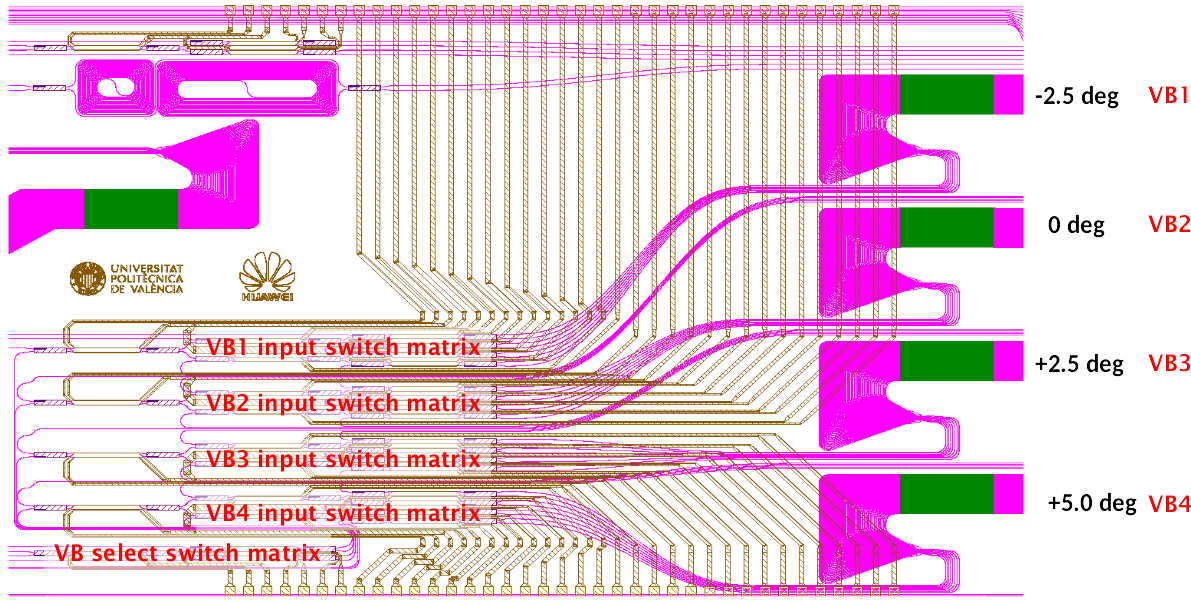}
\includegraphics[width=1\linewidth]{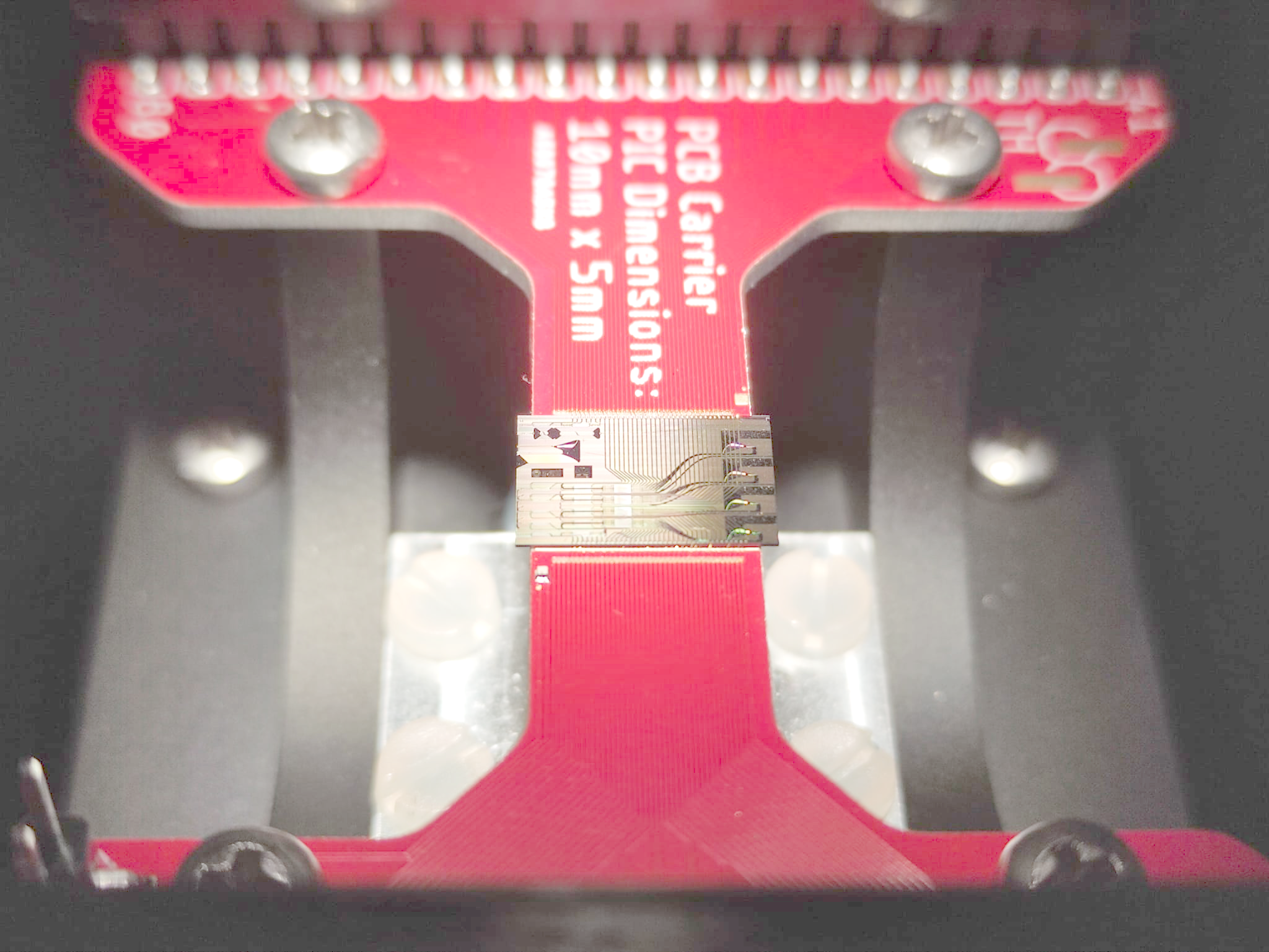}
\caption{SC1 chip layout (top), color legend: purple (waveguides), green (grating couplers), brown (metal tracks). Picture of the PCB mounted chip (bottom).}
\label{fig:sc1_chip}
\end{figure}

\begin{figure}[t]
\centering
\includegraphics[width=1.0\linewidth]{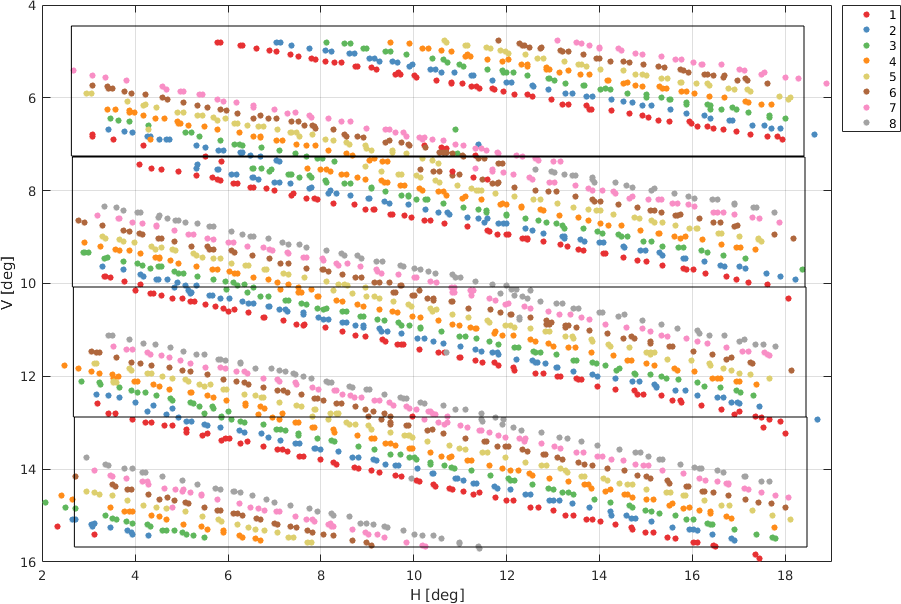}

\caption{SC1 chip device results. Measured peak intensity positions, showing the multi-input interlacing (inputs 1–8, in pink–grey color).
}
\label{fig:sc1_trails}
\end{figure}

The layout and picture of the SC1 chips are given in Fig.~\ref{fig:sc1_chip}. The inputs for each VB OPA are accessed through a switch matrix. A first switch matrix stage allows to select among the different VB OPAs. The beam trajectories for the SC1 chip are given in Fig.~\ref{fig:sc1_trails}. It shows the beam positions for the 4 OPAs each addressing a single VB, comprising a total V FOV of 4~x~2.8\dg, for the TL tuning range of 35~nm. Each VB has been framed by a rectangle within the picture. Notice parts of the FOV are not covered with the present chip, as already discussed in the design section. 

The power of the main peak ultimately determines the detection range and the power in the side lobes is wasted. This is discussed in section~\ref{sec:system} later on. The main lobe power emission efficiency (percentage of power with respect to the input power of the chip) is given in Table~\ref{tab:loss}. The variation range in the whole wavelength tuning range, coming both from intrinsic (wavelength-dependent emission, waveguide width far field effects) or extrinsic (substrate reflection interference effect) can be well appreciated in Fig.~\ref{fig:beams} and are discussed in the following. The SC1 beams were already included in Fig.~\ref{fig:beams}(b) for comparison with SC0. As expected by design, the H beam width is smaller for the SC1 design, since the OPA aperture was doubled. To be precise, the FWHM widths are 0.24\dg~x~0.16\dg (H x V). The V FWHM for both the SC0 and SC1 designs is very similar, as can also be seen visually comparing Fig.~\ref{fig:beams}, since the same GC design was used. The slight differences in the main beam features can also be attributed to the fact SC0 and SC1 were fabricated in different runs at different times. The SC1 V beams exhibit a strong side feature at -0.4\dg from the main lobe, present in most of the max-hold plots (not provided). Compared to the H beam slice where the side lobes appear at different locations, the V side lobe is at a fixed one. Furthermore the latter does not have a fixed intensity. As it can be observed, the overlay of beams fills-up an intensity range from -30 to -3~dB. It might indicate a rapid intensity variation within a short wavelength interval, compatible with some interferometric process from contributions having similar amplitude, in a long cavity created by spurious reflections. As mentioned, noticeable artifacts appear as well in the SC1 H beam, within the same intensity range from peak than for SC0. A thorough investigation on this was not the scope of the proof of concept stage of our research, but we are inclined to believe it's a combination of TE-TM interaction, as we indicated for the SC0 chip, and on-chip reflections.

The experiments were performed operating the switch matrix and the tunable laser. After calibration, the switch matrix bias current required for every state were be stored, and we observed fully reproducible matrix configuration along the experiments. For the results shown, the TL was continuously tuned for every matrix selected SC-OPA input. However there architecture can sport further scanning schemes, some of them we outline hereby. 

\begin{itemize}
\item Random access: the TL wavelength range and switch matrix state can be arbitrary. Thus, parts of the scene can be accessed, i.e. addressing those areas in the FOV where targets have been identified (e.g. fast moving targets) while neglecting others (static or slower moving targets).
\item Adjustable resolution: since the switch matrix state can be arbitrary, two adjacent SC inputs could be enabled at the same time. Two spots would be created simultaneously (i.e. two of the lines shown in the trajectories figures scanning parallel in time), which would represent and effective reduction of the spatial resolution. Time wise, with respect of LiDAR system integration time (see section ahead) the scan rate could be halved to preserve the same SNR, owing to the fact the optical power in this situation would is split in two beams, instead of one.
\item Object framing: two far inputs, e.g. input 1 and 8 could be enabled simultaneously. This way, an outline of a large object, duly identified by processing algorithms in a regular higher resolution spatial scan. \end{itemize}

In summary, the architecture allows for addressing parts of the scene randomly, with different spatial resolutions. Considerations on the laser and switch matrix set and reset times should be addressed, but are out the scope of this paper.

In terms of footprint, the switch matrix (SM) stage, in its most simple structure without resorting to special layouts for the MZIs, scales in length as the number of required stages. That is, if P is the number of inputs to one of our SC-OPAs, the SW stage length will be $L_{SM}=L_{MZI} \log_2(P)$, and the height (or width) will be $W_{SM}=W_{MZI} P/2$. Thus, the total area , where the factor $1/2$ stems from the tree-like structure. In our design, this amounts for 5.14x0.4~\mms (see Fig.~\ref{fig:sc1_chip}), which is 2~\mms area. On other hand, the SC-OPA without switching, that gives the same scene line density, has to scan P times faster for a wavelength sweep $\Delta\lambda_{TL}$. Consequently, the path length increment between adjacent paths from the SC to the grating couplers, $\Delta L$ , must be multiplied by a factor $P$. In general, if we consider a rectangular layout for the array of waveguides between the SC and the grating couplers, the footprint scales $\propto \Delta L^2$. Thus, a length increase factor of $P$ for $\Delta L$ results into an increase of $P^2$ in the used area. In our design, the area used by the arrayed waveguides is approximately 1~\mms (see Figs.~\ref{fig:sc0_chip} and \ref{fig:sc1_chip}). Hence, resorting to a non-switched version of the SC-OPA would result into 64~\mms of area used, as compared to our SC-OPA with $P=8$. In conclusion, the switched architecture presents a footprint ratio advantage, as compared to the non-switched one, of (1+2)/64 approximately. This ratio is enlarged for larger $P$, since the 1~\mms corresponds to the OPA, and 2~\mms to the SM~2x8.

\section{\label{sec:system}LiDAR system perspective}

\subsection{ToF versus FMCW}
Optical beam steering approach and the ranging technique employed can have crossed implications that we summarize in this section. First of all just recall the main differences between the two main approaches for the range determination. Time of flight (ToF) is a incoherent technique so direct detection (DD) is employed. A train of optical pulses is emitted and a broad area receiver based on broad aperture optics focus the incoming beam to the detector plane (broad area or and array of detectors). Finally, the range of the target is obtained by evaluation of the time delay between emitted and received pulse sequence electronically. The Frequency Modulation Continuous Wave (FMCW) technique uses coherent detection. The transmitter emits continuously in power, but modulated in optical frequency with a symmetrical saw-tooth shape \cite{fmcw1}\cite{fmcw2}. A portion of the generated FMCW signal is applied to the detector (as local oscillator, LO) jointly with the target received signal.  The range information will emerge as a beating RF tone between 0~Hz up to the peak to peak optical frequency modulation employed ${\Delta}f$. The FMCW technique is attracting a lot of interest due to these advantages: a) High sensitivity owing to the coherent detection, where almost quantum shot noise limit can be reached; b) interfering signals from other vehicles or from the sun radiation are highly eliminated; c) relative velocity of moving targets respect to the LiDAR can be retrieved because of the Doppler effect from the obtained RF tones due to double ramp saw-tooth design. The target velocity information is being envisioned as a key parameter on the decision taking algorithms for autonomous driving and other LiDAR applications. 

In general FMCW presents a number of advantages versus ToF, but it has an intrinsic requirement that is the coherent beating on the detector of the two combined signals, LO and received signal from target. This in turn can pose restrictions on the practical structure for the TRx/RCx pair. In this sense, on ToF-LiDAR the transmitter and receiver can be separated devices with optimised designs and just located together with only electronic interconnections with the control system. However, in FMCW the LO signal from the transmitter must be driven to the detector for the beating. In free space optics based LiDARs employing large area detectors, or detector arrays, this could be a complex issue taking into account that incoming signals can reach the receiver from a broad field of view (FOV) and the efficient control of the LO and signal overlapping at the detector would require complex solutions. According with this intrinsic requirement, a natural approach is to employ the same optical device for the beam steering for transmission and reception, separating the two propagation directions by means of non-reciprocal components as optical circulators. Those are very mature for the case of discrete optical fiber components, but present very promising evolution as integrated components \cite{circulator}. 

\subsection{Beam steering strategies and ranging techniques}
This section presents some of the implications between scanning strategies and distance detection techniques. First of all we must recall an intrinsic limitation relative to the speed of light. The round trip time for each meter in free space is approximately 6.66~ns. Depending on the application it will determine the minimum "waiting time" (WT) (time between the ranging signals have been emitted and they comeback to the LiDAR receiver). As example, for autonomous driving where the maximum ranges can be $>200$~m waiting time can reach $WT=1.33$~$\mu$s. Additionally to WT, a key parameter is the "processing time" (${\tau}_p$). This quantity includes all the treatments over the detected photo current, such as filtering,  sampling, Fourier transforming depending of the employed approach, and thresholding. All these procedures can be carried out in analogue or digital manner (or as a combination). In an optimum design case, we could assume full advantage of the the complete ${\tau}_p$ is taken for the "signal integration" or in other words for the noise reduction. This is the case for the electrical noise bandwidth $B_e$ on the Signal to Noise Ratio (SNR) models where $B_e=1/{\tau}_p$. Values for ${\tau}_p$ will depend on the maximum range, source power and component losses (we develop them later on) but for AD applications it can range $>1-2$~${\mu}s$. Notice at this point that WT + ${\tau}_p$ is imposing a limitation on the frame rate. For example, a 300x300 complete image will be limited to a 3.7 frames/s assuming an optimistic $WT+{\tau}_p=3$~${\mu}s$. Additionally to WT and the processing time, the establishment and stabilization times must be considered in each LiDAR image point for both steering process (i.e. switching matrix control or MEMs activation time, etc), and also for the ranging signal generation (i.e. FMCW saw-tooth generation and linear frequency control system\cite{linealization}). Several techniques exist for the linearization of a continuously tunable laser. The interested reader may resort to \cite{Zhang:19} and references therein. This evidences a strong requirement for parallel processing in high-performance LiDAR, where multiple system functions must be carried out in parallel (i.e. beam steering along with proper FMCW frequency generation) in order to shorten the total per point time.   

After the previous general considerations we can conclude the beam steering approach and the signals employed for the ranging/speed determination, including their optimum relationships, will determine the final LiDAR features. Thus, we dedicate the rest of the section to explore these relations for the proposed architecture, assuming the use of FMCW ranging signals.

A classical approach is the point to point switching. It implies a sequence of steps: 1) Beam steering and stabilization time. It encompass two actions, the wavelength tuning to perform both the H and V steering into the Vertical Block (VB) and the switching matrix to select the proper VB, 2) The FMCW generation consisting in the saw-tooth frequency modulation with ${\Delta}f$ excursion and 2T period. The theoretical limit for the spatial ranging resolution is only related with ${\Delta}f$ as ${\Delta}R=c/2{\Delta}f$ with c the speed of light. As a reference, in autonomous driving resolutions below 15~cm are demanded, that correspond with ${\Delta}f>1$~GHz. The coarse wavelength tuning for beam steering, and the fine linear optical frequency sweep, are carried out and optimised in a separated way, resulting in greater time consumption for the entire process.
Furthermore, there is a unavoidable cross-relationship between FMCW signals and wavelength tuning steering systems, which is the beam deviation according to the frequency modulation of the saw-tooth signal itself. In these cases, it must be ensured that the angle variation induced by FMCW signal is smaller than the desired beam divergence (or alternatively image point separation). As previously described vertical dispersion on GC was $(\partial\theta/\partial\lambda)_V$=0.08\dg/nm but horizontal dispersion due to the delay line section previous the OPA is $(\partial\phi/\partial\lambda)_H=N_{lines}FOV_H/P\Delta\lambda$, being for $N_{lines}=4$ and $P=1$ a value of $(4x15$\dg/35nm)=1.71\dg/nm.

Notice that in this case, the maximum FMCW frequency excursion $\Delta f$ is 7.2GHz (case when H beam deflection equals the desired H divergence of 0.1 degree). It is important to note that this maximum feasible excursion decreases as the number of lines into the VB increases. For example for $P=1$ and $N_{lines}$ = 28 (compatible with the vertical density of 0.1 degrees along 2.8\dg), the maximum ${\Delta}f$ is 1~GHz (close to a range resolution of 15~cm as previously indicated). In this sense, the proposed switched architecture once again plays to our advantage by decoupling the $N_{lines}$ concept from the $N_{cycles}$ through the product $N_{lines}=P\times N_{cycles}$, increasing the possible FMCW excursion by a factor $\times$ P. 

\begin{figure}[bt]
\centering
\includegraphics[width=1\linewidth]{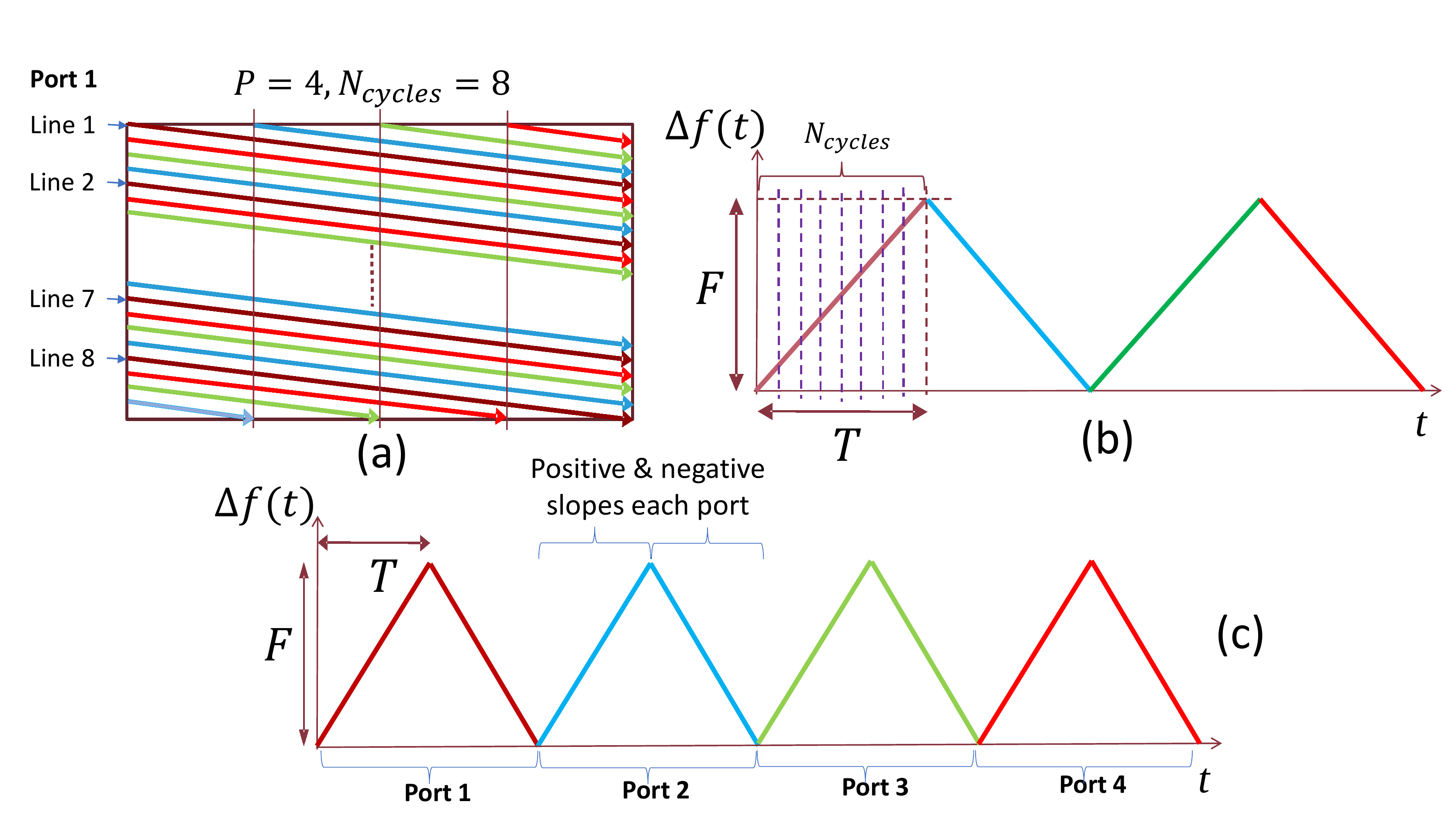}
\caption{Steering trajectories inside a Vertical Block and their optical frequency correspondence. a) Interleaved trajectories. Lines of port 1 are indicated. b) Single ramp per input port. c) Double ramp for relative velocity determination}
\label{fig:FMCW}
\end{figure}

An alternative approach employs the laser sweep to accomplish both the beam steering and the FMCW ranging process simultaneously, in a continuous way \cite{Okano:20}. Fig.~\ref{fig:FMCW}. shows schematically the steering trajectories inside a vertical block and their optical frequency correspondence, to illustrate the connection between steering and FMCW process and their interdependence. In order to simplify the representation, and to make easier the explanation, we take an specific combinations of $P=4$ (number of input ports to the OPA slab coupler (SC)), and $N_{cycles}=8$ (the number of drawn lines along the VB when one complete wavelength scan is accomplished). Notice that 32 oblique lines will fill in full the vertical block field, leading to $2.8º/32=0.088$ vertical interline space. Different colors are employed for each SC input, showing the line steering interleaving into the VB, and their correspondent frequency variation. 
In the standard FMCW technique, both positive and negative frequency slopes must be employed, in order to retrieve the magnitude, and the direction of the target, relative velocity without ambiguity. To achieve this, two consecutive frequency scans with positive and negative slopes must be completed per each input port as depicted in Fig.~\ref{fig:FMCW}(c). We define the $T_{VB}$ as the available time to scan a complete VB. The frame time will be $T_{frame}=T_{VB}N_{VB}$, the time to scan all the lines at each SC input will be $T_{port}=T_{VB}/P$ and finally $T$ in Fig. \ref{fig:FMCW} $T=T_{VB}/2P$ the single ramp time. 
As explained up to this point, the single point classical FMCW approach has been expanded into a continuum swept extended to a certain frequency range (F) that is shared along the whole set of image points covered in each SC port input. In practice, this continuous swept is effectively sliced after detection by the signal processor in a set of "time beams", according with the specified angular resolution and/or beam divergence ($div$). This way, the number of "time beams" into the single time ramp $T$ can be formulated as $M_{FMCW}=\gamma(N_{cycles}FOV_H)/div$, with $FOV_H$ the horizontal field of view and $\gamma$ an over-slicing parameter $\gamma>=1$. Once $M_{FMCW}$ is set, the effective optical frequency excursion is obtained as ${\Delta}f_{eff}=F/M_{FMCW}$, and from this the expected range resolution ${\Delta}R=c/2{\Delta}f_{eff}$ together with the maximum RF tones that should be detected and processed $RF_{tone}={\Delta}f_{eff}$. Table~\ref{tab:table1} summarises the relevant system results for a frame rate equal to 10~Hz, $N_{VB}=10$ ($FOV_{V}=2.8^o \times N_{VB}$) leading to $T_{VB}=10$~ms, $F=35$~nm $\times$ (100GHz/0.8nm), $FOV_{H}=30^o$ and beam divergence $div=0.1$\dg.

Notice that linearized frequency sweep must be accomplished along the whole range F during the time T (see Table~\ref{tab:table1}), which is a challenge for the tunable source as the number of ports (P) increase. Probably the future availability of rapidly tunable hybrid integrated laser sources in ranges of some tens of nanometers \cite{Boller:20,Tran:19} will determine the viability of the FMCW continuous scanning technique in applications with a wide FOV and fast full image rates.

\begin{table}[!t]
\caption{Main LiDAR features vs P and $N_\text{cycles}$\label{tab:table1}}
\centering
\begin{tabular}{|c||c||c||c||c||c|}
\hline
$P$ & $N_\text{cycles}$ & $T$($\mu$s) & $M_\text{FMCW}$ & $\Delta f_\text{eff}$(GHz) & $\Delta R$(m) \\
\hline
1 & 32 & 10000 & 9600 & 0.29 & 0.51\\ \hline
2 & 16 & 5000 & 4800 & 0.58 & 0.26\\ \hline
4 & 8 & 2500 & 2400 & 1.11 &  0.13\\ \hline
8 & 4 & 1250 & 1200 & 2.33 & 0.064\\\hline
16 & 2 & 625 & 600 & 4.66 & 0.032\\ \hline

\end{tabular}
\end{table}

\subsection{Power budget and LiDAR range}
\begin{figure}[bt]
\centering
\includegraphics[width=1\linewidth]{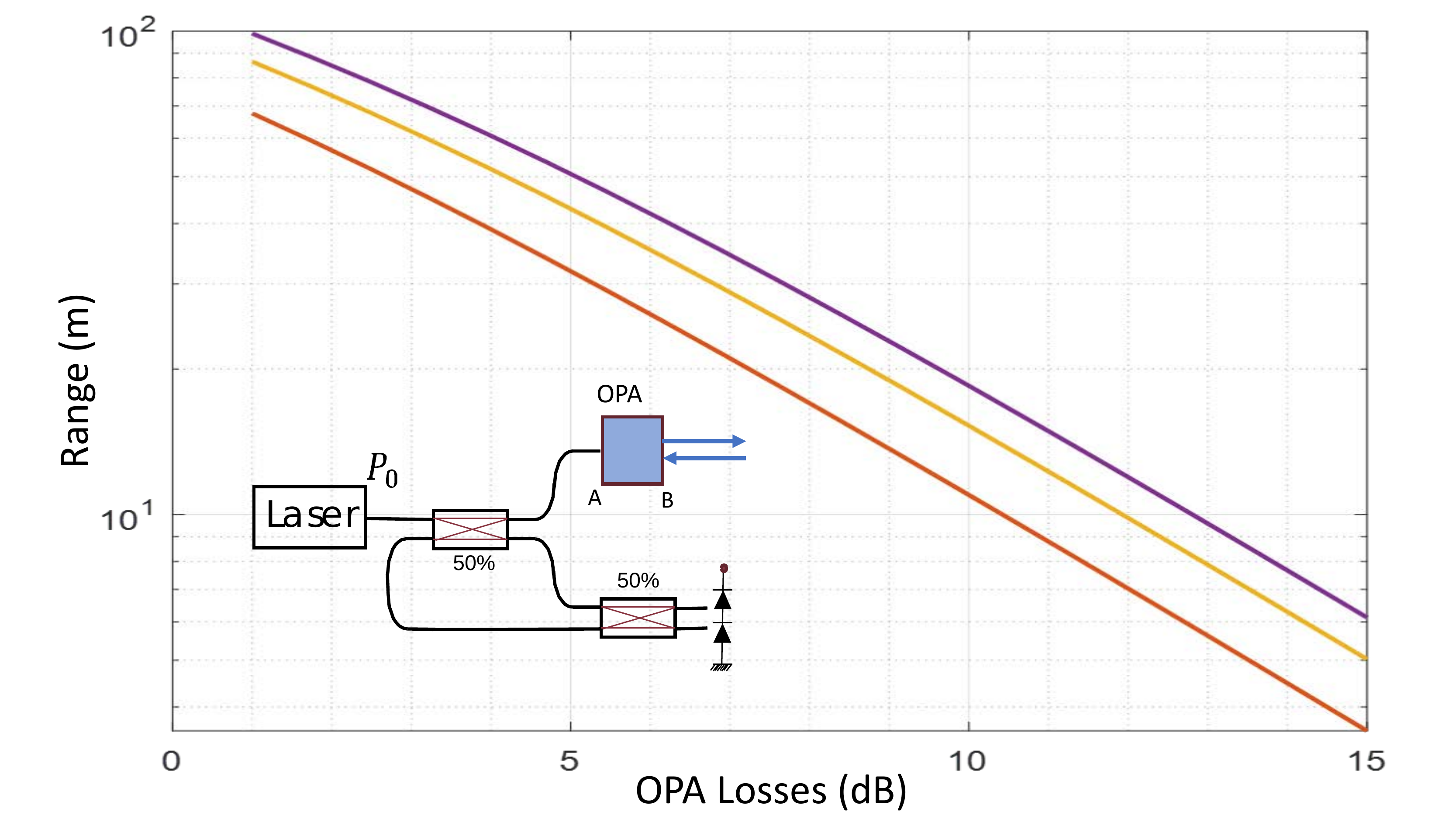}
\caption{Maximum range for the system parameters. Curves for $P_0$ = 100
, 200 and 300 mW.}
\label{fig:Range_Loss}
\end{figure}

A good approximation to determine the LiDAR achievable range is usually based on a shot noise dominant approximation. We will assume the laser source provides a linewidth less than a few tens of kHz, so that beating noise can be neglected against the inherent level of shot noise. The FMCW technique works in a coherent detection regime, applying a relatively high optical power coming directly from the FMCW source that acts as the local oscillator, in order to obtain the RF tone after beating with the received signal. Usually in these cases the thermal noise can be considered below the shot noise, and neglected as well. The inset in Fig.~\ref{fig:Range_Loss} shows a basic LiDAR arrangement with transmission and reception in the same device. We assume a 2$\times$2 (50\%) splitter to divide the FMCW signal to the OPA path and to the detectors path (local oscillator signal). Also a 2$\times$2 (50\%) splitter is employed as combiner before the differential detector pair. The powers considered in Fig.~\ref{fig:Range_Loss} are similar to those in \cite{Li:21}, where 400~mW, from which 100~mW were diverted to a Ge-on-Si photo-detector, and used for ranging over 100~m.

\begin{table}[!t]
\caption{System Passive Components and room for Improvement}
\label{tab:loss}
\centering
\begin{tabular}{|c||c||c|}
\hline
\textbf{Device} & \textbf{Losses(dB)} & \textbf{Improved(dB)} \\
\hline
Fiber/Chip Coupling & 0.5 &  0.4 \\ \hline
2x4 VB selector(2 stages) & 0.46x2x2=1.84 & 0.2x2x2=0.8  \\ \hline
2x8 SC input sel.(3 stages) & 0.46x2x3=2.76 & 0.2x2x3=1.2 \\ \hline
SC insert losses & 4 & 2 \\ \hline
OPA gratings efficiency & 3 / 0.5 & 3 / 0.5 \\ \hline
Grating Apodization & 3 & 0.5\\ \hline
Propagation losses & 0.3 & 0.15 \\ \hline
TOTAL OPA (L$_{AB}$) & 15.4 / 12.9 & 8.05 / 5.55\\ \hline
\end{tabular}
\end{table}


In order to summarize the results of maximum achievable range for the FMCW-LiDAR, we will focus on Fig.~\ref{fig:Range_Loss}, where the maximum range (m) is represented against the total optical losses on the OPA (only one direction in between A and B in the scheme). 
The parameters adopted for the calculation are: "processing time" ${\tau}_p=2$ $\mu$s, receiver aperture area in which the electromagnetic field is picked up onto the LiDAR $A=1cm^2$, the reflectance of the target is 10\% and detectors quantum efficiency $\eta=0.8$. No absorption in the media is considered, being the back propagation losses determined by the fractional area between the receiving aperture area (A) over the isotropic reflected field on target as $l_{prop}=10log_{10}(A/4\pi {L_{target}}^2)$
Finally, it is necessary to set the probability of target detection PD=90\% and the probability of false alarm PFA$<10^{-4}$. PD is the probability of detection in presence of the target, and PFA is the probability of detection when no target is in the image. PFA occurs when the shot noise level exceeds the established electronic detection threshold on the detected RF tones. At the same time the threshold level is settled once the noise power is estimated (dominated by shot in our case) and according to the established maximum admissible PFA. Note that an average of 25 false alarm events will occur in a single second according to the pixels rate of 250 Mpixel/s ($1/(2{\tau}_p)$), which may be surprising, but it must be borne in mind that this is the first level of processing and that other higher processing levels can rule out many of these random events \cite{Threshold_PD_PFA_1}\cite{Threshold_PD_PFA_2}. 

The maximum range results are shown, for three values of optical power injected by the laser source P$_0$=100, 200 and 300~mW. Clearly, the available laser power and OPA losses are extremely demanding parameters over the maximum range, so that, even assuming an optimistic value of 6~dB for the OPA losses, distances of 26 meters are obtained for 100 mW, which can be increased to approximately to 43 m for 300~mW. 

OPA losses $L_{AB}$ are summarized in a detailed manner in Table~\ref{tab:loss}. The second column shows the results extracted from measurements along the technology and BB characterization runs. The third column presents an estimate of the achievable loss values, assuming a foreseeable evolution and improvement of the technological processes.
Apart from the well-known losses in the switch matrices (in total 5 stages of MZIs are counted), we can summarize some of the other sources of losses and their possible improvement: 1) Slab Coupler losses are dominated by the Gaussian profile truncation and the efficiency on field coupling to the output guides \cite{Munoz:02}. 2) OPA grating efficiency is mainly related with 50\% loss of power due to the bottom radiation on symmetrical designs. Vertical symmetry breaking approaches \cite{Raval:17} can reduce the expected loss up to few tenths of a dB. 3) The apodized design of the GC is essential in reducing the Main to Secondary Lobe Ratio up to expected goal on designs ($MSLR>30dB$). In addition, its final length and the apodization profile also determine the width of the far-field beam divergence goal ($div=0.1^o$). For these values, the total required length for the Gaussian near-field profile was 1005.5~nm. The next step was to define the local perturbation along the GC that produces the expected near-field profile taking in mind that the signal applied propagating along the GC is being reduced progressively (due to waveguide losses and mainly radiated fraction of the light) and therefore the perturbation depth profile, $de(z)$, will be a distorted version of the initial Gaussian goal. This can be easily done by recursive numerical calculations up to obtain the optimum $de(z)$. Notice that, once MSLR and $div$ are set, for a given maximum achievable perturbation depth ($de_{max}$) the optimum design provides a maximum ratio of total radiated power respect to the applied one (value on Table~\ref{tab:table1} as grating apodization). In our design $de_{max}$ was set to 115~nm that corresponds with a maximum $\alpha_{GC-TE}$=8~dB/mm local power decaying due grating radiation, leading to a total grating apodization losses of 2.34~dB when the complete profile is evaluated. This grating apodization losses can be reduced by increasing the maximum achievable $\alpha_{GC-TE}$ through increasing $de_{max}$ and/or the mode profile overlapping with the waveguide perturbed cross-section. Moreover, longer devices allow the reduction of the angular divergence ($div$) but also the losses. As an example, a 2~mm long GC with equivalent apodization profile and same $de_{max}$ as designed would result into a 0.83~dB losses. 4) Propagation loss measured was 0.183~dB/cm and reports \cite{Pfeiffer:18} show results as good as 0.055~dB/cm, considering interconnection waveguides represent a total length of 1.5~cm.

Starting from the indicated base system, it is interesting to have simple relationships that allow for evaluating the impact on the maximum achievable range by simple calculations. More specifically, each dB increase or decrease in OPA losses affects the range by a factor $\times$0.81 or $\times$(1/0.81) respectively, while in optical power P$_0$ each dB impacts by a factor $\times$1.11 or $\times$(1/1.11). Finally, a $\times$2 increase in the OPA aperture A results in a $\times$1.358 range increase, while in the case of the process time T a factor $\times$2 produces an improvement in the range of $\times$1.433.

\subsection{Switch matrix cross-talk impact}
The MZI based limited extinction for the 1xP~SM preceeding the SC-OPA, can lead to the emission of P-1 spurious beams evenly spaced in the horizontal FOV by FOV$_H$/P, with intensities -20~dB (worst case, and the closest to the main radiated beam). -23~dB for 2~beams, -26~dB for 4~beams, and so on.

In transmit / receive mode, the optical signals (main beam and spurious) go through the OPA and SM twice, hence the cross-talk ratio is a factor of x2. In the worst case, the cross-talk of a spurious to the main beam would then be -40~dB. Considering this worst case, two different ranging and detection scenarios may arise:
\begin{enumerate}
    \item Same distance to target than to spurious direction. In that case, after beating with the LO FMCW signal, two RF tones of the same frequency are obtained, which add in photocurrent. The amplitude for the interferer in photocurrent will be 100 times lower (beat term square root, $\sqrt{10^4}$) thus with negligible effect over the main beam signal.
    \item Spurious beam comes from an object located closer or further from the main beam target. In this case, after the detector, there are 2 RF tones, main and spurious, but with different RF frequency. Thus, they can be distinguished and separated by processing the received signals. Even though, their value will be very low as explained in the following.
\end{enumerate}

The worst case of the two outlined in the last point above, would be when the spurious beam comes from an object closer ($L_{sp}$) and with reflectivity higher, to the main beam target ($\rho_{sp}$). The power ratio between spurious and main beam can be written as: $X_t(dB) = -40 + 10\log_{10}(\rho_{sp}/\rho_{main}) + 20\log_{10}(L_{main}/L_{sp})$. The first term accounts for the different reflectivity, and the second from the different range (see propagation loss equation in the manuscript).

As numerical example, for $\rho_{sp}/\rho_{main}$=10 and $L_{main}/L_{sp}$=10, $Xt = -10$~dB (10 in natural units). In that case, the spurious would be detected with an amplitude level of $\sqrt(10)$=3.16 but a frequency 10~times lower.

In conclusion, simple mechanisms could be established for the processing of the detected signal, that would allow to discard tones below a given threshold from the main beam tone. In any case, those resulting from a SM with cross-talk -20~dB, can be neglected.

\subsection{Power amplification}
\begin{figure}[bt]
\centering
\includegraphics[width=1\linewidth]{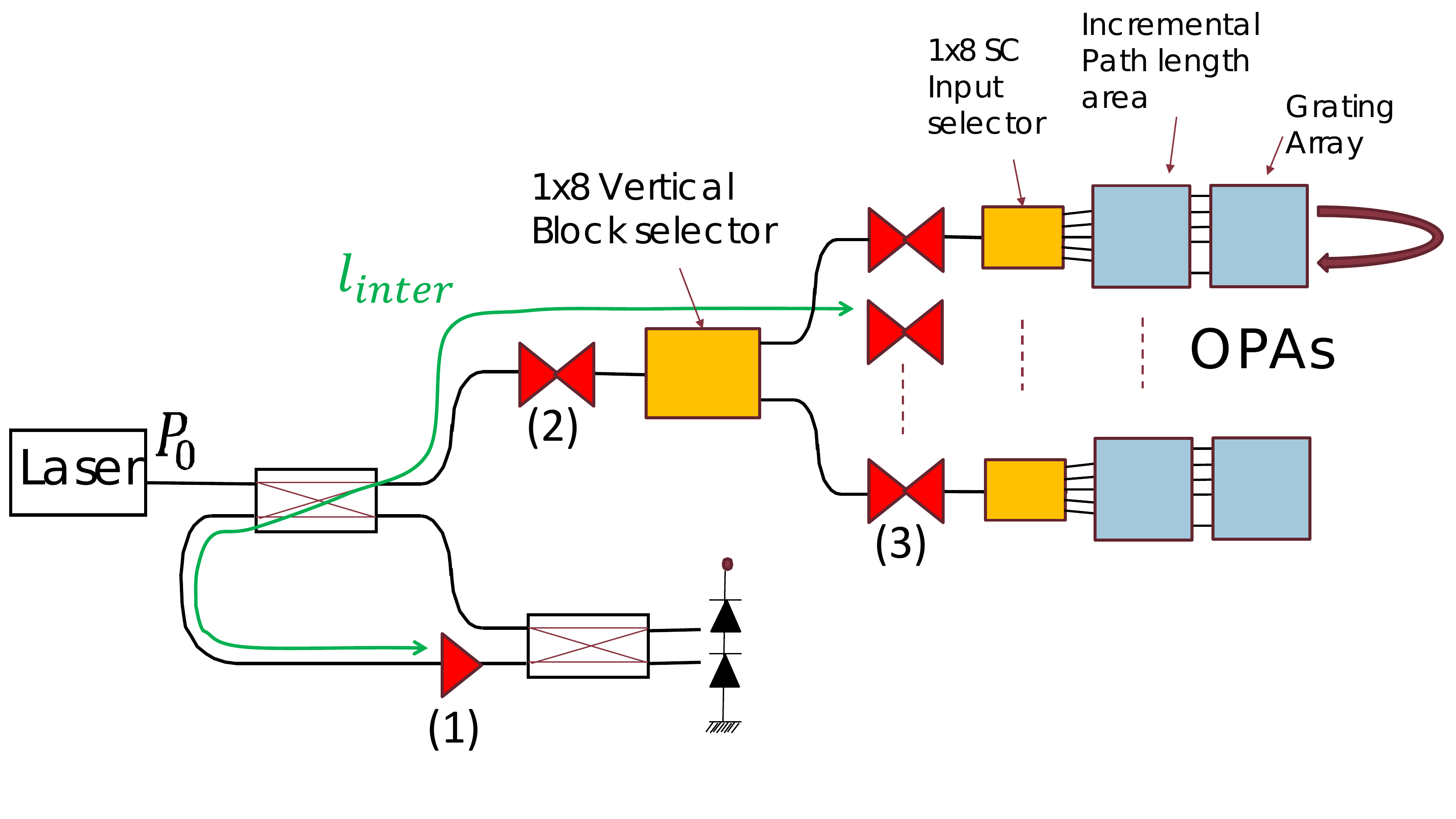}
\caption{Optical amplification. Three OA locations are considered. $l_{inter}$ are the one pass losses in between the OA and the reciber input. }
\label{fig:amplifiers}
\end{figure}

We consider here the benefits of using optical amplifiers (OA) embedded into the specific points of the LiDAR architecture. We begin placing the OA just before the signal input port of the differential detector in the typically called pre-amplifier configuration (case on Fig.~\ref{fig:amplifiers} location (1)). As can be envisioned this approach results into the addition of a new noise term coming from beating noise between the local oscillator (LO) and the amplified spontaneous emission noise (ASE) from OA. This noise term is given by ${\sigma^2}_{LO-ASE}=2{\eta}^2e^2P_{LO}NB_{e}$ where $\eta$ is the detectors quantum efficiency, ${e}$ is the electron charge, $P_{LO}$ is the LO power at the detector input, $B_{e}$ the electrical bandwidth and $N=n_{sp}(G-1)$ with $n_{sp}$ the OA population inversion factor and ${G}$ the gain. On the other hand, the shot noise term is ${\sigma^2}_{shot}=2e({\eta}e/h{\nu})P_{LO}B_{e}$ and comparing the two expressions we have ${\sigma^2}_{LO-ASE}/{\sigma^2}_{shot}={\eta}n_{sp}(G-1)$. We can see that pre-amplier configuration leads to a dominant LO-ASE term approximately ${G}$ times the shot noise, preventing any hypothetical benefit assumed from optical gain.

Lets consider now a more general configuration where the OA is located far from the detector input, so that a certain quantity of optical losses are in between (case on Fig.~\ref{fig:amplifiers} locations (2) and (3))). We  employ ${l_{inter}<1}$ to insert this into the formulation, so the noise ratio remains as ${\sigma^2}_{LO-ASE}/{\sigma^2}_{shot}={\eta}n_{sp}(G-1){l_{inter}}$. We can define a signal to noise ratio at the receiver output in the two cases: a) limited by shot noise (no OA): $SNR_{0}=S_0/{\sigma^2}_{shot}$ and b) Employing OA:$SNR={S_0{G^m}}/{({\sigma^2}_{shot}+{\sigma^2}_{LO-ASE})}$ and finally the ratio between them that provides the net gain benefit: $G_{net}=SNR/SNR_{0}=G^m/(1+{\eta}n_{sp}(G-1)l_{inter})$. 
Notice that we have assumed in this generalised approach that the OA can be localised in different places along the LiDAR architecture, where the signal bi-directionality must me guaranty (cases 2 and 3). In these cases the optical gain is assumed equal in the two directions, and the double pass benefit over the signal is taken into account by the ${m=2}$ parameter on the formulas.

We will illustrate the OA net gain benefits in three different locations along the LiDAR architecture. For all the cases we take a modest gain G=4 (6~dB), $NF=2n_{sp}$ and ${\eta}=0.8$. 1) AO just before detectors (location (1) in Fig.~\ref{fig:amplifiers}) so $l_{inter}=1$ (no intermediate losses) and the net gain is $G_{net}=0.81$ (-1dB). 2) The amplifier is in position (2) in Fig. \ref{fig:amplifiers}, in between the ideal 2$\times$2 splitter acting as duplexor and the OPAs. In this case $l_{inter}=1/2$ (3dB) and just evaluating the net gain with ${m=2}$ we obtain $G_{net}=6.66$ (+8.23dB). 3) 8 amplifiers are located at each output of the Vertical Block switching matrix where the switching matrix is supposed to have in total 3~dB excess losses. In this case we apply $l_{inter}=1/4$ and the net gain is $G_{net}=13.33$ (+11.24~dB). Notice that each 2~dB obtained in $G_{net}$ must be counted as an effective reduction of 1~dB in the OPA losses, leading to an increase of $\times(1/0.81)$ in range. In the previous cases of positions 2 and 3, the range increase should be $\times$2.34 and $\times$3.27. 

In summary, amplification on chip is anticipated as a must. Furthermore, a distributed set of amplifiers with modest gain are outlined as the optimum in terms of signal to noise ratio.

\section{\label{sec:concl}Conclusion}
A fully passive two-dimensional OPA based on wavelength sweeping and consisting of a multi-input star coupler has been proposed and demonstrated in silicon nitride technology. Existing slab coupler OPA permits to distribute optical power between the numerous radiating elements in a single step and creating a Gaussian profile in the H direction. Our novel proposal uses $P$ inputs at the entrance of the slab coupler, that allows to relax the up-scaling problem with the star coupler approach, permitting to reduce both footprint and optical power losses in a $P$ factor with respect to the single-input approach. Expanded results for a previously reported proof-of-concept device have been reported, featuring a H$\times$V FOV of 15\dg$\times$2.8\dg, and a beam size of 0.36\dg$\times$0.175\dg, for wavelength sweep of $\Delta\lambda=35$ nm, as expected by design. A new design incorporating on-chip switch matrix stages to select among four different vertical blocks, features H$\times$V FOV of 15\dg$\times$11.2\dg, and a beam size of 0.24\dg$\times$0.167\dg. The architecture can be operated as progressive scene scanner, but it can also feature random scene access with selectable resolution.

From a LiDAR system perspective, we have shown the proposed OPA is compatible with FMCW LiDAR, taking advantage of the tunable laser wavelength sweep. Moreover the switched architecture has the ability to reduce by a $P$ factor the horizontal angular dispersion, for a given line density, which in turn reduces the impact of the FMCW signals on the steering deviation. Additionally, the relationship between FMCW frequency tuning in continuous mode with the interlaced steering trajectories has been described in detail, as well as the implication of the double slope use for the target relative speed determination. Dependency relationships between the steering architecture parameters $P$ and $N_{cycles}$ and system parameters of interest, such as the number of FMCW points, the equivalent per point frequency excursion and the range resolution have been discussed. 

The power balance and distance limitations have been described, providing quick-use scaling rules that can be quickly used for the evaluation of maximum-range with respect to losses. A breakdown of the current losses and paths for improvements have been provided for the proposed architecture and technology. Finally the use of optical gain inside the LiDAR structure is theoretically evaluated, in what we believe may represent a path of disruptive change toward a new generation of LiDAR systems that will build upon hybrid photonic integration.

\section*{Acknowledgment}
We acknowledge Ligentec, the silicon nitride technology platform company responsible for the fabrication of the chip measured in this work. Equipment and infrastructure funding by GVA/IDIFEDER/2018/031.

\ifCLASSOPTIONcaptionsoff
  \newpage
\fi


\newpage

\begin{IEEEbiography}[{\includegraphics[width=1in,height=1.25in,clip,keepaspectratio]{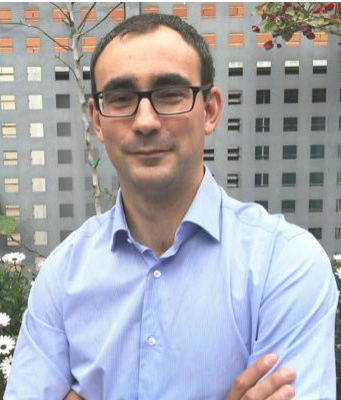}}]{Pascual Mu\~noz} (S'98, M'04, SM'13) was born in Valencia, Spain on February 7th, 1975. He  received the Ingeniero de Telecomunicaci\'on degree from Universitat Polit\`ecnica de Val\`encia (UPV) in 1998. In 1999 he served as 1st Lieutenant in the Spanish Airforce, while working as IT consultant for AIME Instituto Tecnol\'ogico. He received a Ph.D. degree in photonics from UPV in 2003. He is currently Full Professor at the Departamento de Comunicaciones, and researcher at the Institute for Telecommunications and Multimedia Applications (ITEAM), both at UPV.
Dr. Mu\~noz runs a consolidated research line, started in 2005, on prototyping Photonic Integrated Circuits (PICs) in a technology agnostic fashion, where PICs are designed in the best suited technology (Silicon-On-Insulator, Indium Phosphide, Silica on Silicon, Silicon Nitride amongst other) for each application. He has been involved in several European Commission funded projects, being coordinator for integration on InP within the NoE IST-EPIXNET. He has published over 50 papers in international refereed journals and over 80 conference contributions.  He is a member of the Technical Programme Committees of the European Conference on Optical Communications (ECOC) and the European Conference on Integrated Optics (ECIO). 
Dr. Mu\~noz received the VPI Speed Up Photonics Award in 2002 for innovative Fourier optics AWG with multimode interference (MMI) couplers modeling, by Virtual Photonics Incorporated and IEEE Communications Magazine. He was also granted the IEEE/LEOS Graduate Student Fellowship Program in 2002. He received the extraordinary doctorate prize from UPV in 2006. From his research line, he co-founded the UPV spin-off company VLC Photonics in 2011, where the PIC design know-how, expertise and tools have been transferred, and he served as CEO from 2011 to 2013. Dr. Mu\~noz is a Senior Member of the IEEE and a Senior Member of the OSA.
\end{IEEEbiography}

\begin{IEEEbiography}[{\includegraphics[width=1in,height=1.25in,clip,keepaspectratio]{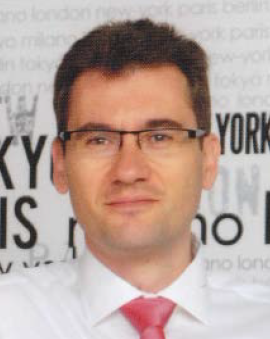}}]{Daniel Pastor} was born in Elda, Spain, in 1969. His research career at photonics technologies started at 1993 when joined the Optical Communication Group as researcher and the Communication Department at Universitat Politècnica de València (UPV) as lecturer and PhD student. From 1994 to 1998 he was a lecturer at the Telecommunications Engineering Faculty and he became an Associate Professor in 1999. He obtained a PhD in Electrical Engineering and Telecommunications by the Universitat Politècnica de València, Spain in 1996, with Thesis topic “Redes de Difracción sobre Fibra Óptica para Procesado de Señales Fotónicas”. He is co-author of more than 225 papers in journals and international conferences in the fields of Optical Delay Line Filters, Fiber Bragg Gratings, Microwave Photonics, WDM and SCM lightwave systems, Optical Fiber Sensors, Silicon Nitride integrated devices for telecom and sensing, and advanced characterization techniques as Optical Frequency Domain Reflectrometry. His current technical interests include Photonic Integrated Circuits for telecom, LiDAR and chemical and bio-sensing applications.
\end{IEEEbiography}

\begin{IEEEbiography}[{\includegraphics[width=1in,height=1.25in,clip,keepaspectratio]{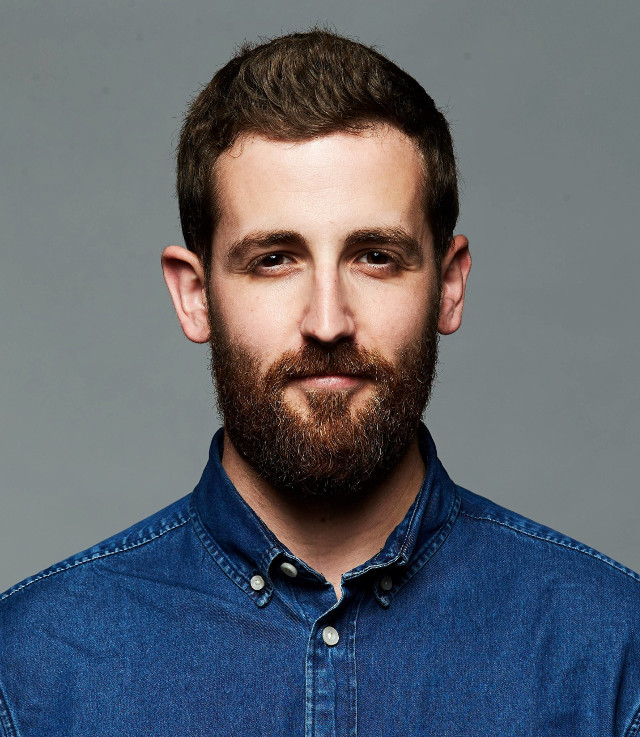}}]{Luis A. Bru} was born in Elche, Spain, in 1988. He received the B.Sc degree in Physics in 2012, and the M.Sc degree in Advanced Physics (specialty on Photonics) in 2013, both at the Universitat de València (UV), Valencia, Spain. He is currently a post-doc researcher at Photonics Research Labs (PRL), Universitat Politècnica de València (UPV), where he has been enrolled from 2015 to date. In 2022, he received a Ph.D. degree in Photonics, where Optical Frequency Domain Interferometry (OFDI) technique is explored in the context of photonic integrated circuits.
With more than 25 co-authored publications in scientific journals and conferences, he has been involved in research projects including PIC topics such as silicon nitride integrated devices, integrated LiDAR, or hybrid photonic integration.
Beyond, his research interests expand also to quantum mechanics and quantum information, having kept a collaboration with nonlinear and quantum optics research group at UV from 2013 to 2017.
\end{IEEEbiography}

\begin{IEEEbiography}[{\includegraphics[width=1in,height=1.25in,clip,keepaspectratio]{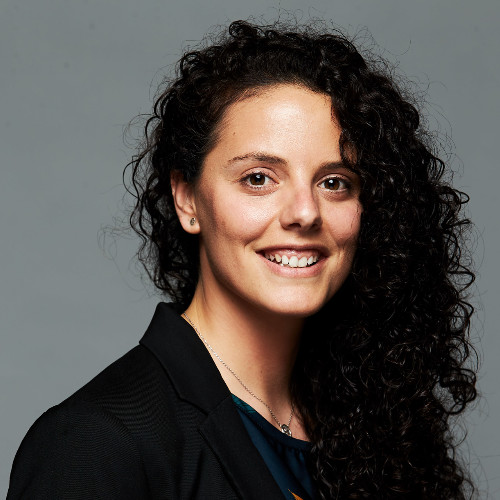}}]{Gloria Micó} graduated with a Bachelor of Physics in 2011 at the University of Valencia (UV). She obtained her M.Sc. in Photonics from the Polytechnic University of Catalonia in 2013, developing her Master thesis on biophotonics at Institute of Photonics Sciences (ICFO). During 2015-2020 she developed her Ph.D. thesis focused on the design and characterization of photonic integrated spectroscopic sensors and their applications at Photonics Research Lab (PRL) at the Universitat Politècnica de València (UPV). Since 2019, she has been working on projects for private companies related to LIDAR technology. Nowadays, she is working at UPVfab as facility manager. 
\end{IEEEbiography}

\begin{IEEEbiography}[{\includegraphics[width=1in,height=1.25in,clip,keepaspectratio]{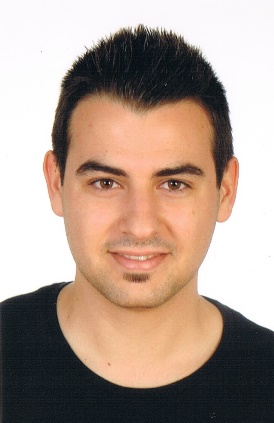}}]{Jesus Benítez} obtained his Bachelor's degree in Telecommunications Engineering from the Universitat Politècnica de València (UPV) in 2015. In the same year, he joined the Photonics Research Labs (PRL) from UPV as a P.hD. student. During 2015-2019, he developed his P.hD. in Microwave Photonics applied to Low Coherence Interferometry. From 2019-2021 he has been collaborating in private funded projects related to photonic integrated circuits (PIC) as a post-doctoral researcher. Currently, he is performing lab operations, testing and system validation in iPronics Programmable Photonics. His interests cover PIC characterization and microwave photonics.
\end{IEEEbiography}

\begin{IEEEbiography}[{\includegraphics[width=1in,height=1.25in,clip,keepaspectratio]{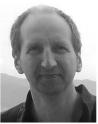}}]{Dominic Goodwill} received the Ph.D.degree in physics from Heriot-Watt University, Edinburgh, U.K., in 1991.He has been a Senior Principal Engineering the Advanced Photonics Team of Huawei Canada since 2012. He is focused on data center and transport applications of silicon photonics, and optical technologies for 5G wireless networks. From 2004 to 2011, he was an applications software architect for Nortel and Genband corporations, creating IP television solutions, and real-time serverapplications deployed to millions of consumers. From 1997 to 2004,he led Nortel’s optical interconnects technology team, creating free-space optics and a world-first 10GigE DWDM photonic switched Metro network. He led the fiber optics standard for the Infiniband trade association. He was chair for IEEE Optical Interconnects conference, and a program committee member for Optical Fiber Communication conference. Previously, at Heriot-Watt University and University of Colorado, he researched III-V semiconductors and polymers for optical computing and interconnect. He has more than40 issued patents

\end{IEEEbiography}

\begin{IEEEbiography}[]{Eric Bernier} 
\end{IEEEbiography}

\end{document}